\documentclass[lettersize,journal]{IEEEtran}
\usepackage{amsmath,amsfonts}
\usepackage{algorithm}
\usepackage{array}
\usepackage[caption=false,font=normalsize,labelfont=sf,textfont=sf]{subfig}
\usepackage{textcomp}
\usepackage{stfloats}
\usepackage{url}
\usepackage{verbatim}
\usepackage{graphicx}
\usepackage{cite}
\usepackage{algpseudocode} 
\usepackage{multicol} 
\usepackage{placeins} 
\PassOptionsToPackage{hidelinks}{hyperref} 
\usepackage{orcidlink} 

\begin{document}

\title{A Lightweight Cubature Kalman Filter for Attitude and Heading Reference Systems Using Simplified Prediction Equations}

\author{Shunsei Yamagishi\orcidlink{0000-0003-2480-3185}, Lei Jing\orcidlink{0000-0002-1181-2536}, \IEEEmembership{Member, IEEE}
\thanks{This work was supported by NEDO Intensive Support for Young Promising Researchers Number 21502121-0, Collaborative Research with Toyota Motor Corporation, and JKA and its promotion funds from KEIRIN RACE.}
\thanks{Shunsei Yamagishi and Lei Jing are with Graduate School of Computer Science, Engineering, The University of Aizu, Aizuwakamatsu 965-8580, Japan (e-mail: d8271107@u-aizu.ac.jp; leijing@u-aizu.ac.jp).}}

\maketitle

\begin{abstract}
Attitude and Heading Reference Systems (AHRSs) are broadly applied wherever reliable orientation and motion sensing is required. In this paper, we present an improved Cubature Kalman Filter (CKF) with lower computational cost while maintaining estimation accuracy, which is named "Kaisoku Cubature Kalman Filter (KCKF)". The computationally efficient equations of the KCKF are derived by simplifying those of the CKF, while preserving equivalent mathematical relations. The lightweight prediction equations in the KCKF are derived by expanding the summation terms in the CKF and simplifying the result. This paper shows that the KCKF requires fewer floating-point operations (FLOPs) than the CKF. The controlled experimental results show that the KCKF reduces the computation time by approximately 19\% compared to the CKF on a high-performance computer, whereas the KCKF reduces the computation time by approximately 15\% compared to the CKF on a low-cost single-board computer. In addition, the KCKF maintains the attitude estimation accuracy of the CKF.
\end{abstract}


\section{Introduction}
\IEEEPARstart{T}{he} Attitude and Heading Reference Systems (AHRSs) have been widely investigated in previous studies, since AHRSs are widely used in various industrial fields. AHRSs are integrated into various devices such as robotics, aircraft, automobiles, and smartphones. In addition, AHRSs are widely used in research fields, including navigation systems such as Pedestrian Dead Reckoning (PDR) \cite{b16,b17,b28}, motion capture systems such as human joint angle estimations \cite{b26,b23,b35,b36}, and attitude estimation algorithms.\par
One of the research topics in AHRSs and Magnetic and Inertial Measurement Units (MIMUs) is attitude estimation, and many papers have investigated various attitude estimation algorithms. The attitude estimation algorithms are generally based on Wahba's problem \cite{b32}, Kalman filter \cite{b1,b2,b7,b8,b9,b10,b13,b14,b15,b16,b17,b22,b23,b25,b28,b29}, Complementary filter (CF) \cite{b5,b6,b24,b27}, gradient descent method \cite{b4,b11, b12, b13, b30}. Moreover, Kalman filter-based approaches are divided into categories of the Linear Kalman filter (LKF), Extended Kalman filter (EKF), Unscented Kalman filter (UKF), and the Cubature Kalman filter (CKF).\par
The CF estimates the attitude by combining the weighted estimate from the angular velocity and the weighted estimate from accelerometer and magnetometer measurements in the frequency domain. J. Wu et al. \cite{b6} proposed the Fast Complementary Filter (FCF), which is faster than Madgwick's algorithm \cite{b4} and the LKF by avoiding iterative computations. H. Rong et al. \cite{b24} proposed CF, which avoids the effect of magnetic disturbance on roll and pitch angle estimations. Additionally, H. Rong et al. \cite{b27} proposed a time-efficient complementary Kalman gain filter (TCF) that achieves accuracy close to that of the EKF, while having a lower computational cost by avoiding the computation of the inverse matrix required for the TCF gain calculation.\par
The gradient descent-based algorithms update the attitude by minimizing the error function. The Madgwick filter \cite{b4} estimates the attitude by fusing a quaternion updated by a gradient-descent method and a quaternion obtained by integrating angular velocity, using a complementary-filter-based approach. K. Sever et al. \cite{b30} proposed a CF based on the gradient-descent method that reduces the required number of gradient-descent iterations.\par
The Kalman filter-based approaches update the attitude by the statistical model. The Kalman filter-based approaches were researched by \cite{b1,b2,b7,b8,b9,b10,b13,b14,b15,b16,b17,b22,b23,b25,b28,b29}. S. Guo et al. \cite{b1} proposed the fast quaternion-based Kalman filter (FKF), which is faster than existing Kalman filters and complementary filters, while maintaining accuracy. The Guo's FKF achieves low computational cost by using low cost observation formulas \cite{b1}. L. Xue et al. \cite{b22} researched the Kalman filter approach for fusing two Micro Inertial Measurement Units (MIMUs). L. Xue et al. \cite{b22} showed that their proposed method has lower attitude errors than a single MIMU. Some observation vectors of the Kalman filter are derived from Wahba's problem \cite{b32}. X. Chen et al. \cite{b25} and L.-F. Shi et al. \cite{b28} proposed the observation models that fuse quaternions calculated by the QUaternion ESTimator (QUEST) \cite{b33} and the Factored Quaternion Algorithm (FQA) \cite{b34}. By fusing QUEST and FQA in the observation model, the accuracy is improved over the Kalman filter with an observation model only using either QUEST or FQA \cite{b25}.\par
The EKF-based approaches are popularly used to estimate the attitude of the AHRS. The EKF can update the state vector using the non-linear state model and observation model. The EKF linearizes the nonlinear state and observation models in the current estimate to first order using a Taylor expansion \cite{b14}. The observation model of the EKF was simplified by Z. Dai et al.\cite{b2}. The observation vector of the Lightweight Extended Kalman Filter (LEKF) proposed by Z. Dai et al. \cite{b2} requires only four elements of the Direction Cosine Matrix (DCM), and the LEKF achieves a lower computational cost. X. Wei et al. \cite{b29} proposed the Measurement Adaptive Reduced Direction Cosine Matrix (MA-RDCM), which is applied to the EKF and has higher accuracy than existing methods under dynamic conditions.\par
However, there are cases that the accuracy of the EKF is not enough, since the EKF simply approximates the non-linear model to the first-order of the Taylor expansion. The UKF can compute higher accuracy estimates than the EKF \cite{b14}. The UKF handles nonlinear models by generating $2N_q+1$ sigma points, where $N_q$ is the dimension of the state vector, and achieves higher accuracy than the first-order approximation of the EKF. The UKF was proposed by S. J. Julier and J. K. Uhlmann \cite{b21}. The UKF based approaches for estimating the attitude were researched by \cite{b7}, \cite{b8}, \cite{b10}, \cite{b17}. S. Yamagishi and L. Jing \cite{b17} reduced the computational cost of the UKF for estimating the attitude of the AHRS by simplifying the Unscented Transform formulas.\par
Nevertheless, the UKF has a higher computational cost than the EKF \cite{b17}. More recently, the Cubature Kalman filter was proposed by I. Arasaratnam and S. Haykin \cite{b19}. The CKF for attitude estimation was researched by \cite{b14}, \cite{b15}. The CKF has also higher accuracy than the EKF \cite{b14}. The CKF handles nonlinear models by generating $2N_q$ cubature points, and achieves higher accuracy than the first-order approximation of the EKF. In addition, there are other advantages in the CKF. According to F. Yu et al. \cite{b18}, the CKF has lower computational cost than the UKF. Furthermore, according to Y. jun Yu et al. \cite[p. 1447]{b15}, “CKF guarantees the positive definiteness of the variance matrix”.\par
However, the CKF still has a higher computational cost than the EKF because of its extensive matrix computations, which hinders its real-time applications like motion capture systems and PDR. In addition, it is important not only to reduce computational cost but also to maintain estimation accuracy. Reducing computational cost will contribute to estimating the attitude on a low-cost computer and to reducing the energy consumption of a smart phone.\par

To this end, in this paper, the CKF with a lower computational cost is proposed. We refer to the proposed CKF as “Kaisoku Cubature Kalman Filter (KCKF)”. “Kaisoku” is a Japanese word, and it means nearly the same as “rapid”. KCKF can compute the attitude of AHRS, while maintaining the accuracy.\par
The contributions of this paper are as follows.
\begin{itemize}
    \item A computationally efficient CKF, named KCKF, is proposed, which can compute the attitude of the AHRS while maintaining the accuracy of the CKF.
    \item The prediction formulas of the KCKF are theoretically derived from those of CKF, keeping the equivalent relation.
    \item The computational cost of the proposed KCKF is compared with the CKF from a theoretical point of view. The computational cost of the KCKF and the CKF is compared by the Floating-Point Operations (FLOPs).
    \item The accuracy and computational cost of the KCKF and the CKF were compared in the controlled experiments.
\end{itemize}

\section{Methods}
\subsection{State Model and Observation Model of Attitude Estimation Algorithm}
In general, attitude estimation algorithms are designed using the state model and the observation model. In this paper, the state model and the observation model described in the papers \cite{b2}, \cite{b9} are applied to the CKF and the proposed KCKF. The state model of the attitude estimation algorithms is defined by (\ref{eq:general_state_model}).
\begin{equation}
    \label{eq:general_state_model}
   \boldsymbol{q}_k=\biggl(\boldsymbol{I}_{4\times4}+\frac{\Delta t}{2}[\boldsymbol{\Omega}_k\times]\biggr)\boldsymbol{q}_{k-1}+\boldsymbol{G}_k\boldsymbol{w}_k
\end{equation}
$[\boldsymbol{\Omega}_k\times]$ is denoted by (\ref{eq:Omega_k}).
\begin{equation}
    \label{eq:Omega_k}
    [\boldsymbol{\Omega}_k\times]=
    \begin{bmatrix}
    0 & -\omega_{x,k} & -\omega_{y,k} & -\omega_{z,k} \\
    \omega_{x,k} & 0 & \omega_{z,k} & -\omega_{y,k} \\
    \omega_{y,k} & -\omega_{z,k} & 0 & \omega_{x,k} \\
    \omega_{z,k} & \omega_{y,k} & -\omega_{x,k} & 0
    \end{bmatrix}
\end{equation}
In this paper, $\boldsymbol{F}_k$ is denoted by (\ref{eq:Fk}).
\begin{equation}
    \label{eq:Fk}
    \boldsymbol{F}_k=\boldsymbol{I}_{4\times4}+\frac{\Delta t}{2}[\boldsymbol{\Omega}_k\times]
\end{equation}
$\begin{bmatrix}
    \omega_{x,k} & \omega_{y,k} & \omega_{z,k}
\end{bmatrix}^T$
is the angular velocity vector measured by the gyro sensor. 
In addition, $\boldsymbol{G}_k$ is denoted by the following formula.
\begin{equation}
    \boldsymbol{G}_k=\frac{\Delta t}{2}
    \begin{bmatrix}
        q_{1,k} & q_{2,k} & q_{3,k}\\
        q_{0,k} & q_{3,k} & -q_{2,k}\\
        -q_{3,k} & -q_{0,k} & q_{1,k}\\
        q_{2,k} & -q_{1,k} & -q_{0,k}
    \end{bmatrix}
\end{equation}

$\boldsymbol{I}_{n\times n}$ denotes the identity matrix of size $n\times n$. Let $\boldsymbol{q}_k=
\begin{bmatrix}
    q_{0,k} & q_{1,k} & q_{2,k} & q_{3,k}
\end{bmatrix}^T$ be a quaternion
, and let $\Delta t$ be the sampling period. The term $\boldsymbol{G}_k\boldsymbol{w}_k$ denotes system noise that follows a normal distribution with $\mathcal{N}(\boldsymbol{0}_{4\times1},\boldsymbol{Q}_k)$, where $\boldsymbol{0}_{4\times1} \text{ and }\boldsymbol{Q}_k$ denote the four-dimensional zero vector and the covariance matrix, respectively. In addition, $\boldsymbol{w}_k\sim\mathcal{N}(\boldsymbol{0}_{3\times1},\sigma^2_{\omega}\boldsymbol{I}_{3\times3}), \boldsymbol{E}[\boldsymbol{G}_k\boldsymbol{w}_k(\boldsymbol{G}_k\boldsymbol{w}_k)^T]=\boldsymbol{Q}_k, \text{ and }\boldsymbol{E}[\boldsymbol{w}_k\boldsymbol{w}_k^T]=\sigma_{\omega}^2\boldsymbol{I}_{3\times3}$, where $\boldsymbol{E}[\cdot]$ denotes the expectation operator. $\boldsymbol{Q}_k$ is calculated using (\ref{eq:Qk}).
\begin{equation}
    \label{eq:Qk}\boldsymbol{Q}_k=\boldsymbol{G}_k(\sigma_{\omega}^2\boldsymbol{I}_{3\times3})\boldsymbol{G}_k^T
\end{equation}
$\sigma_{\omega}^2$ denotes the variance in noise from the gyro sensor.
In this paper, the function $\boldsymbol{f}$ of the state transition is defined by (\ref{eq:state_transition}).
\begin{equation}
    \label{eq:state_transition}
    \boldsymbol{f}(\boldsymbol{q}_k)=\biggl(\boldsymbol{I}_{4\times4}+\frac{\Delta t}{2}[\boldsymbol{\Omega}_k\times]\biggr)\boldsymbol{q}_{k}
\end{equation}
Then, the observation model described in the papers \cite{b2}, \cite{b9} is described. The observation model to estimate the attitude of an AHRS is defined by the following formula \cite{b2}, \cite{b9}.
\begin{equation}
    \label{eq:observation model}
    \boldsymbol{z}_k=\boldsymbol{h}(\boldsymbol{q}_k)+\boldsymbol{v}_k
\end{equation}
$\boldsymbol{h}(\boldsymbol{q}_k)$ is defined by (\ref{eq:def_h(q)}) \cite{b2}, \cite{b9}.

\begin{equation}
    \label{eq:def_h(q)}
    \boldsymbol{h}(\boldsymbol{q}_k)=
    \begin{bmatrix}
        \boldsymbol{C}^S_G(\boldsymbol{q}_k) & \boldsymbol{0}_{3\times3}\\
        \boldsymbol{0}_{3\times3} & \boldsymbol{C}^S_G(\boldsymbol{q}_k)
    \end{bmatrix}
    \begin{bmatrix}
        \boldsymbol{a}^G_{r}\\
        \boldsymbol{m}^G_{r,k}
    \end{bmatrix}
\end{equation}
$\boldsymbol{v}_k$ is observation noise that follows a normal distribution with $\mathcal{N}(\boldsymbol{0}_{6\times1},\boldsymbol{R}_k)$, where $\boldsymbol{0}_{6\times1}, \boldsymbol{R}_k$ denote the six-dimensional zero vector and the covariance matrix, respectively. $\boldsymbol{R}_k$ is calculated by (\ref{eq:Rk}).
\begin{equation}
    \label{eq:Rk}
    \boldsymbol{R}_k=
    \begin{bmatrix}
        \sigma_{acc}^2\boldsymbol{I}_{3\times3} & \boldsymbol{0}_{3\times3}\\
        \boldsymbol{0}_{3\times3} & \sigma_{mag}^2\boldsymbol{I}_{3\times3}
    \end{bmatrix}
\end{equation}
$\sigma_{acc}^2, \sigma_{mag}^2$ denote the variance of noise from the acceleration sensor, and that from the magnetic sensor, respectively. Here, $\boldsymbol{E}[\boldsymbol{v}_k\boldsymbol{v}_k^T]=\boldsymbol{R}_k$. $\boldsymbol{z}_k$ is the observation vector defined by $\boldsymbol{z}_k=
\begin{bmatrix}
    a_{x,k} & a_{y,k} & a_{z,k} & m_{x,k} & m_{y,k} & m_{z,k}    
\end{bmatrix}^T$. 
$
\begin{bmatrix}
    a_{x,k} & a_{y,k} & a_{z,k}    
\end{bmatrix}^T$, 
$
\begin{bmatrix}
    m_{x,k} & m_{y,k} & m_{z,k}    
\end{bmatrix}^T$
denote the normalized acceleration vector measured by the acceleration sensor, and normalized magnetic vector measured by the magnetic sensor, respectively.
$\boldsymbol{C}^S_G(\boldsymbol{q}_k)$ is the Direction Cosine Matrix (DCM) of quaternion which converts from the global coordinate system to the sensor coordinate system. DCM $\boldsymbol{C}^S_G(\boldsymbol{q}_k)$ is defined by (\ref{eq:DCM}). $\boldsymbol{0}_{3\times3}$ denotes the zero matrix of size $3\times3$.

\begin{figure*}
\begin{equation}
    \label{eq:DCM}
    \boldsymbol{C}^S_G(\boldsymbol{q}_k)=
    \begin{bmatrix}
        q_{0,k}^2+q_{1,k}^2-q_{2,k}^2-q_{3,k}^2 & 2(q_{1,k}q_{2,k}+q_{0,k}q_{3,k}) & 2(q_{1,k}q_{3,k}-q_{0,k}q_{2,k})\\
        2(q_{1,k}q_{2,k}-q_{0,k}q_{3,k}) & q_{0,k}^2-q_{1,k}^2+q_{2,k}^2-q_{3,k}^2 & 2(q_{2,k}q_{3,k}+q_{0,k}q_{1,k})\\
        2(q_{1,k}q_{3,k}+q_{0,k}q_{2,k}) & 2(q_{2,k}q_{3,k}-q_{0,k}q_{1,k}) & q_{0,k}^2-q_{1,k}^2-q_{2,k}^2+q_{3,k}^2
    \end{bmatrix}
\end{equation}
\end{figure*}

$\boldsymbol{a}^G_r$ denotes the reference acceleration vector in the global coordinate system, and $\boldsymbol{a}^G_r$ is set to 
$
\begin{bmatrix}
    0 & 0 & 1
\end{bmatrix}^T
$. $\boldsymbol{m}^G_{r,k}$ denotes the reference magnetism reference vector in the global coordinate system. $\boldsymbol{m}^G_{r,k}$ is set by (\ref{eq:ref_mag}) \cite{b6}.
\begin{align}
    \label{eq:ref_mag}
    \boldsymbol{m}^G_{r,k}=
    \begin{bmatrix}
        m_{N,k}\\
        0\\
        m_{D,k}
    \end{bmatrix}
\end{align}

$m_{D,k}, m_{N,k}$ are defined by the following equations \cite{b6}.
\begin{align}
    m_{D,k}&=a_{x,k}m_{x,k}+a_{y,k}m_{y,k}+a_{z,k}m_{z,k}\\
    m_{N,k}&=\sqrt{1-m_{D,k}^2}
\end{align}
By substituting  
$\boldsymbol{a}^G_r=
\begin{bmatrix}
    0 & 0 & 1
\end{bmatrix}^T$, (\ref{eq:ref_mag}), and (\ref{eq:DCM}) into (\ref{eq:def_h(q)}), we obtain (\ref{eq:h(q)}).
\begin{figure*}
\begin{align}
    \label{eq:h(q)}
    \boldsymbol{h}(\boldsymbol{q}_k)=
    \begin{bmatrix}
        2(q_{1,k}q_{3_k}-q_{0,k}q_{2,k}) \\
        2(q_{2,k}q_{3,k}+q_{0,k}q_{1,k}) \\
        (q_{0,k}^2-q_{1,k}^2-q_{2,k}^2+q_{3,k}^2) \\
        (q_{0,k}^2+q_{1,k}^2-q_{2,k}^2-q_{3,k}^2)m_{N,k}+2(q_{1,k}q_{3,k}-q_{0,k}q_{2,k})m_{D,k} \\
        2(q_{1,k}q_{2,k}-q_{0,k}q_{3,k})m_{N,k}+2(q_{2,k}q_{3,k}+q_{0,k}q_{1,k})m_{D,k} \\
        2(q_{1,k}q_{3,k}+q_{0,k}q_{2,k})m_{N,k}+(q_{0,k}^2-q_{1,k}^2-q_{2,k}^2+q_{3,k}^2)m_{D,k}
    \end{bmatrix}
\end{align}
\end{figure*}

\subsection{Kaisoku Cubature Kalman Filter and Cubature Kalman Filter}
The CKF was proposed by I. Arasaratnam and S. Haykin \cite{b19}. The CKF has the advantages of providing better accuracy than the EKF \cite{b14}, lower computational cost compared to the UKF\cite{b18}, and ensuring the positive definiteness of the variance matrix \cite{b15}. However, the CKF has the disadvantage that the CKF has higher computational cost, compared to the EKF. Therefore, in this paper, the KCKF with lower computational cost is proposed.\par
\begin{figure*}[htbp]
  \centering
  \includegraphics[width=140mm]{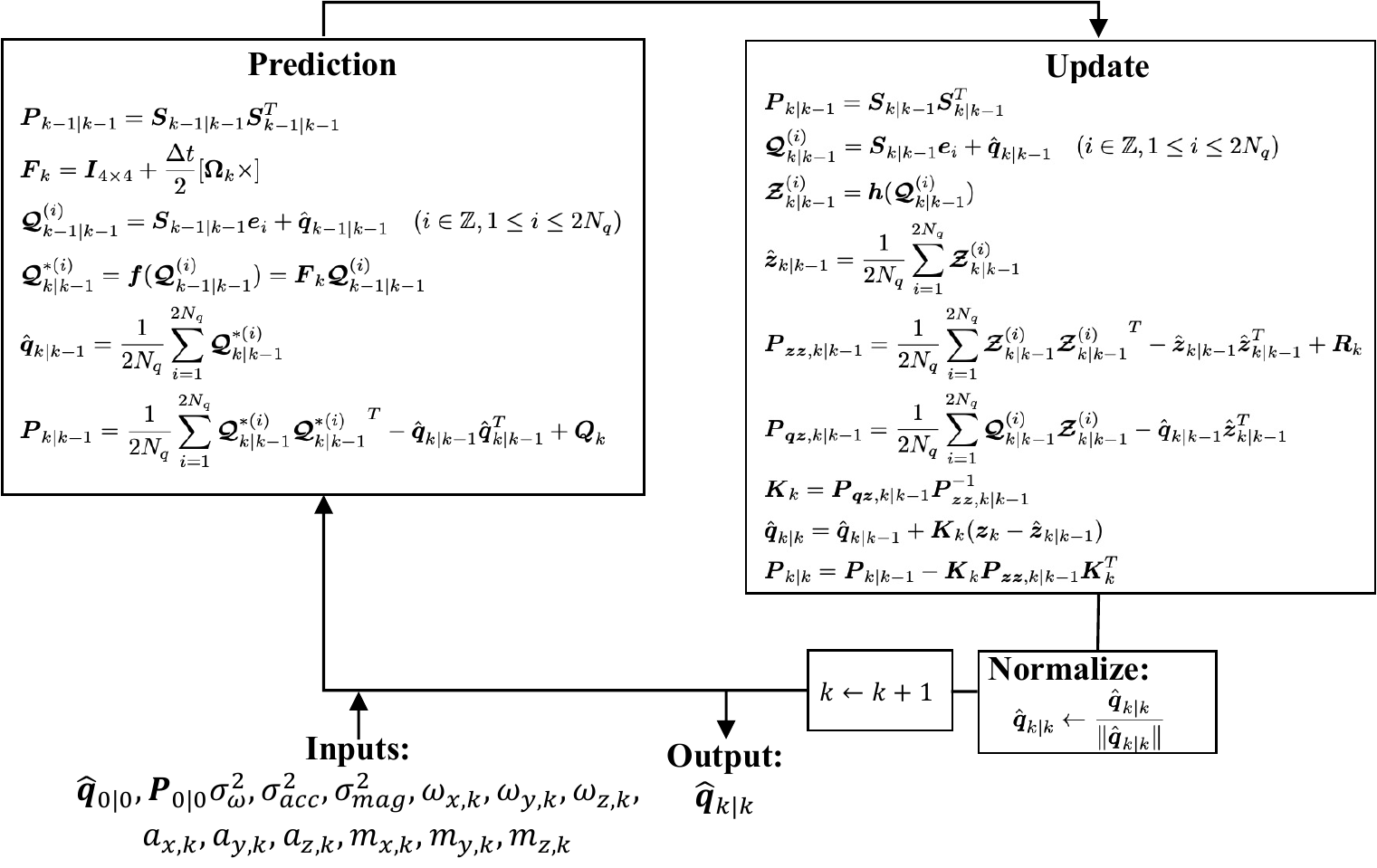}
  \caption{The algorithm of the CKF}
  \label{fig:CKF}
\end{figure*}
\begin{figure*}[htbp]
  \centering
  \includegraphics[width=140mm]{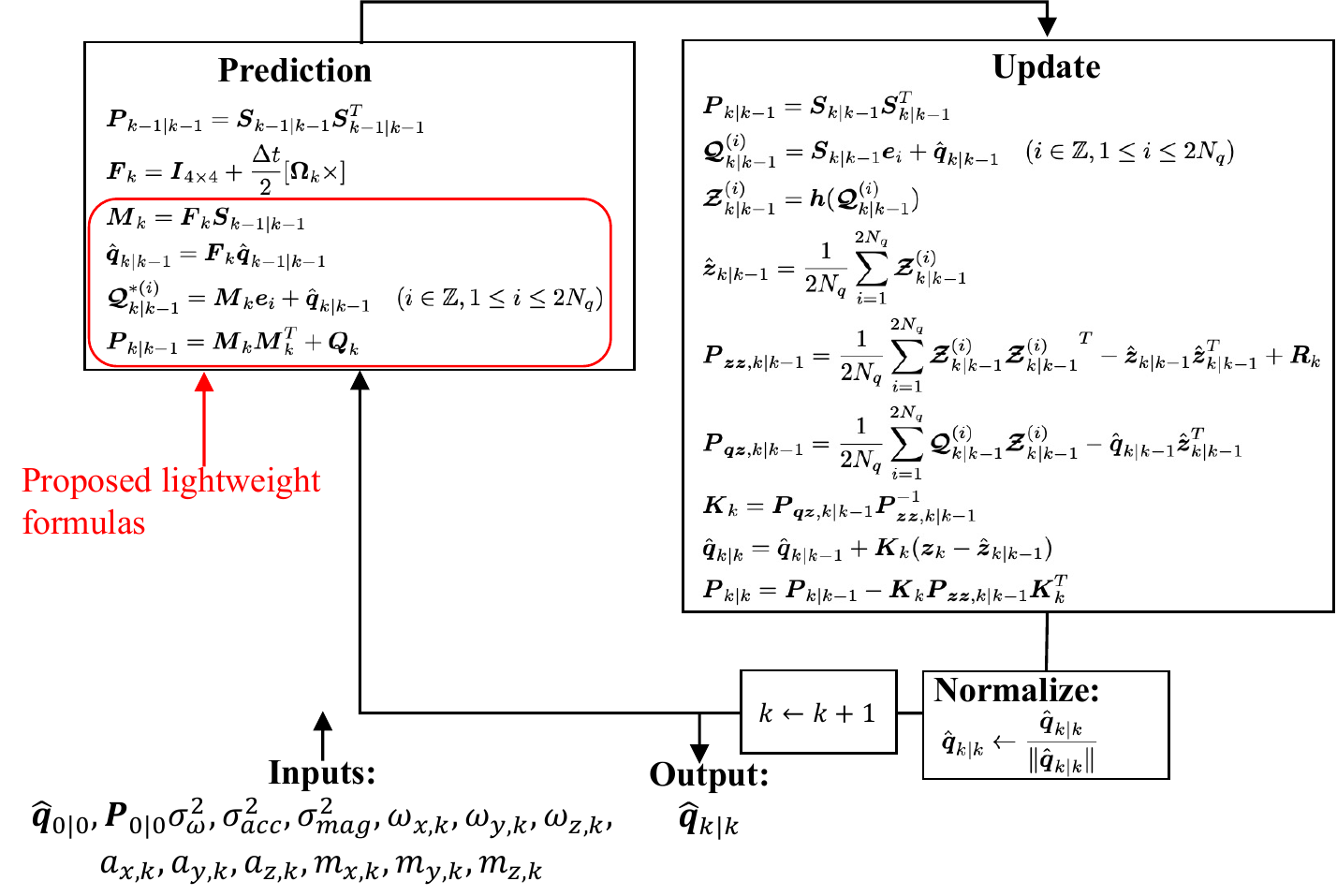}
  \caption{The algorithm of the KCKF}
  \label{fig:KCKF}
\end{figure*}

Figure \ref{fig:CKF} shows the algorithm of the CKF for attitude estimation, while Figure \ref{fig:KCKF} shows the algorithm of the KCKF.

$\boldsymbol{\mathcal{Q}}^{(i)}_{k-1|k-1}, \boldsymbol{\mathcal{Q}}^{*(i)}_{k|k-1}, \boldsymbol{\mathcal{Z}}^{(i)}_{k|k-1}$ denote the cubature points, propagated cubature points in prediction, and propagated cubature points in observation, respectively. $\boldsymbol{e}_i$ is defined by $\boldsymbol{e}_i=\sqrt{N_q}[\boldsymbol{1}]_i$, where $N_q$ denotes the dimension of quaternion. Thus, $N_q$ is set to four. $[\boldsymbol{1}]_i$ denotes $i$th vector in the set $[\boldsymbol{1}]$. $[\boldsymbol{1}]$ is defined by the following formula.
\begin{align}
    &[\boldsymbol{1}]=\{
    \begin{bmatrix}
        1\\
        0\\
        0\\
        0\\
    \end{bmatrix},
    \begin{bmatrix}
        0\\
        1\\
        0\\
        0\\
    \end{bmatrix},
    \begin{bmatrix}
        0\\
        0\\
        1\\
        0\\
    \end{bmatrix},
    \begin{bmatrix}
        0\\
        0\\
        0\\
        1\\
    \end{bmatrix},
    \begin{bmatrix}
        -1\\
        0\\
        0\\
        0\\
    \end{bmatrix},
    \begin{bmatrix}
        0\\
        -1\\
        0\\
        0\\
    \end{bmatrix},
    \begin{bmatrix}
        0\\
        0\\
        -1\\
        0\\
    \end{bmatrix},
    \begin{bmatrix}
        0\\
        0\\
        0\\
        -1\\
    \end{bmatrix}
    \}
\end{align}

The differences between the CKF and the proposed KCKF are in the prediction equations. The prediction equations of the KCKF have lower computational cost than those of the CKF.
The following are the proposed formulas of the KCKF.
\begin{align}
    \label{eq:KCKF1}
    \boldsymbol{M}_k=\boldsymbol{F}_k\boldsymbol{S}_{k-1|k-1}\\
    \label{eq:KCKF2}
    \hat{\boldsymbol{q}}_{k|k-1}=\boldsymbol{F}_k\hat{\boldsymbol{q}}_{k-1|k-1}\\
    \label{eq:KCKF3}
    \boldsymbol{\mathcal{Q}}^{*(i)}_{k|k-1}=\boldsymbol{M}_k\boldsymbol{e}_i+\hat{\boldsymbol{q}}_{k|k-1}\\
    \label{eq:KCKF4}
    \boldsymbol{P}_{k|k-1}=\boldsymbol{M}_k\boldsymbol{M}_k^T+\boldsymbol{Q}_k
\end{align}
On the other hand, the following formulas of the CKF correspond to the proposed formulas of the KCKF.
\begin{align}
    \label{eq:CKF1}
    &\boldsymbol{\mathcal{Q}}^{(i)}_{k-1|k-1}=\boldsymbol{S}_{k-1|k-1}\boldsymbol{e}_i+\hat{\boldsymbol{q}}_{k-1|k-1}\\
    \label{eq:CKF2}
    &\boldsymbol{\mathcal{Q}}^{*(i)}_{k|k-1}=\boldsymbol{f}(\boldsymbol{\mathcal{Q}}^{(i)}_{k-1|k-1})=\boldsymbol{F}_k\boldsymbol{\mathcal{Q}}^{(i)}_{k-1|k-1}\\
    \label{eq:CKF3}
    &\hat{\boldsymbol{q}}_{k|k-1}=\frac{1}{2N_q}\sum_{i=1}^{2N_q}\boldsymbol{\mathcal{Q}}^{*(i)}_{k|k-1}\\
    \label{eq:CKF4}
    &\boldsymbol{P}_{k|k-1}=\frac{1}{2N_q}\sum_{i=1}^{2N_q}\boldsymbol{\mathcal{Q}}^{*(i)}_{k|k-1}{\boldsymbol{\mathcal{Q}}^{*(i)}_{k|k-1}}^T-\hat{\boldsymbol{q}}_{k|k-1}\hat{\boldsymbol{q}}_{k|k-1}^T+\boldsymbol{Q}_k
\end{align}
The formulas of the KCKF from (\ref{eq:KCKF1}) to (\ref{eq:KCKF4}) provide the same computational results as the formulas of the CKF from (\ref{eq:CKF1}) to (\ref{eq:CKF4}), since there is an equivalent relation between them. 

\subsection{Derivation of the Prediction Formulas of the KCKF}
The prediction equations of the KCKF are derived by similar derivation methods to those of the Kaisoku Unscented Kalman Filter (KUKF) \cite{b17}. The main points of the derivations of the prediction equations of the KCKF can be summarized as follows.
\begin{itemize}
    \item (\ref{eq:KCKF2}) is derived by expanding the summation in (\ref{eq:CKF3}) and simplifying the result. 
    \item (\ref{eq:KCKF3}) is derived from (\ref{eq:CKF2}) by substituting (\ref{eq:CKF1}) into (\ref{eq:CKF2}). Then, (\ref{eq:KCKF1}), (\ref{eq:KCKF2}) are used to derive (\ref{eq:KCKF3}).
    \item (\ref{eq:KCKF4}) is derived by expanding the summation in (\ref{eq:CKF4}) and simplifying the result. After expansion, (\ref{eq:KCKF4}) is derived using (\ref{eq:KCKF1}) and (\ref{eq:KCKF2}).
\end{itemize}
For more details, full derivations are provided in “Appendix A POOF”.

\subsection{Comparing the Computational Cost of KCKF With That of CKF}
In this section, the computational cost of the KCKF is compared with that of the CKF theoretically. As in \cite{b17}, floating-point operations (FLOPs) are used to compare computational costs. We adopt the definition of FLOPs introduced in \cite{b37}. An addition, a subtraction, or a multiplication between real numbers is counted as 1 FLOP, and a division between real numbers is counted as 4 FLOPs \cite{b37}. In this paper, we assume that the matrix products are computed according to the pseudo-code below.
\begin{algorithm}[tb]
\caption{Matrix Product Between $\boldsymbol{A}\in\boldsymbol{R}^{l\times m}$ and $\boldsymbol{R}^{m\times n}$}
\label{alg-max-from-array}
\begin{algorithmic}[1]
\Function {Matrix\_Product}{$\boldsymbol{A},\boldsymbol{B}$}
    \For {$i=1$ to $l$} 
        \For{$j=1$ to $n$}
            \State $c_{ij}=\sum\limits_{k=1}^m a_{ik}b_{kj}$
        \EndFor
    \EndFor
    \State \Return $\boldsymbol{C}\in\boldsymbol{R}^{l\times n}$
\EndFunction
\end{algorithmic}
\end{algorithm}
To compute the matrix product $\boldsymbol{C}=\boldsymbol{A}\boldsymbol{B}$, $l\times m \times n$ FLOPs for the multiplications of real numbers and $l\times(m-1)\times n$ FLOPs for the summations of real numbers are needed, where $\boldsymbol{C}\in\boldsymbol{R}^{l\times n},\boldsymbol{A}\in\boldsymbol{R}^{l\times m},\boldsymbol{B}\in\boldsymbol{R}^{m\times n}$.\par
In this comparison, we assume that $\boldsymbol{Q}_k$ is already calculated by (\ref{eq:Qk}).\par
First, the FLOPs of the CKF are considered. To compute a (\ref{eq:CKF1}), 16 FLOPs for multiplications of real numbers and 16 FLOPs for summations of real numbers are needed. Thus, the $2N_q$ equations in (\ref{eq:CKF1}) require 128 FLOPs for multiplications of real numbers and 128 FLOPs for summations of real numbers. To compute a (\ref{eq:CKF2}), 16 FLOPs for multiplications of real numbers and 12 FLOPs for summations of real numbers are needed. Thus, the $2N_q$ equations in (\ref{eq:CKF2}) require 128 FLOPs for multiplications of real numbers and 96 FLOPs for summations of real numbers. In addition, to compute (\ref{eq:CKF3}), 5 FLOPs for multiplications, 4 FLOPs for division of real numbers and 28 FLOPs for summations of real numbers are needed. To compute (\ref{eq:CKF4}), 161 FLOPs for multiplications, 4 FLOPs for division of real numbers and 144 FLOPs for summations of real numbers are needed. In total, 826 FLOPs are needed to compute all the formulas from (\ref{eq:CKF1}) to (\ref{eq:CKF4})\par
Then, FLOPs of the KCKF are considered. To compute (\ref{eq:KCKF1}), 64 FLOPs for multiplications of real numbers, 48 FLOPs for summations of real numbers are needed. To compute (\ref{eq:KCKF2}), 16 FLOPs for multiplications of real numbers and 12 FLOPs for summations of real numbers are needed. Furthermore, to compute a (\ref{eq:KCKF3}), 16 FLOPs for multiplications of real numbers and 16 FLOPs for summations of real numbers are needed. Thus, the $2N_q$ equations in (\ref{eq:KCKF3}) require 128 FLOPs for multiplications of real numbers and 128 FLOPs for summations of real numbers. To compute (\ref{eq:KCKF4}), 64 FLOPs for multiplications of real numbers and 64 FLOPs for summations of real numbers are needed. In total, 524 FLOPs are needed to compute all the formulas from (\ref{eq:KCKF1}) to (\ref{eq:KCKF4}).\par
Therefore, the FLOPs of the KCKF are lower than the FLOPs of the CKF, and the proposed KCKF has lower computational cost than the CKF. Reducing FLOPs will contribute to estimating the attitude on a low-cost computer and to reducing the energy consumption of a smart phone.

\section{Experiments}
\begin{figure}[htbp]
  \centering
  \includegraphics[width=80mm]{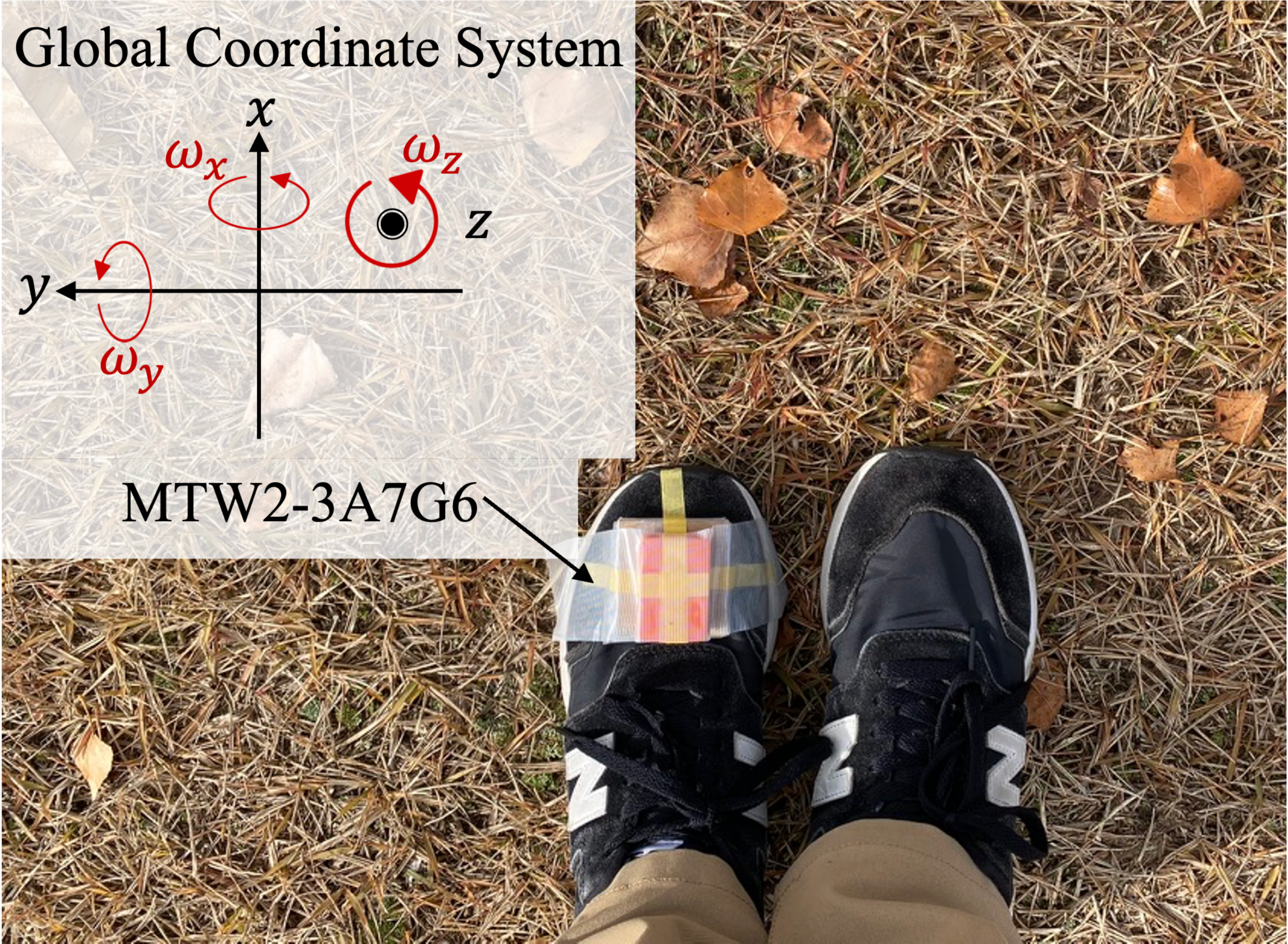}
  \caption{MTW2-3A7G6 mounted on a shoe}
  \label{fig:AHRS}
\end{figure}

In this paper, two experiments were conducted to compare the KCKF with the existing Kalman filters, using the same experimental data as in \cite{b17}. The MTW2-3A7G6 \cite{b20} produced by Movella Inc. was used to collect AHRS data\cite{b17}. All the data were collected outdoors \cite{b17}. Table \ref{table:Xsens_performance} shows the performance of MTW2-3A7G6, and Table \ref{table:data_details} shows the list of data collected in \cite{b17}. Figure \ref{fig:AHRS} shows the mounting place of MTW2-3A7G6 and the definition of the sensor coordinate system.\par
The sampling rate of MTW2-3A7G6 was set to 100Hz \cite{b17}. The parameters of all the Kalman filters were set to $\hat{\boldsymbol{q}}_{0|0}=
\begin{bmatrix}
    1 & 0 & 0 &0
\end{bmatrix}^T,
\boldsymbol{P}_{0|0}=\boldsymbol{I}_{4\times4},\sigma_{\omega}^2=1.000\times10^{-3},\sigma_{acc}^2=1.000\times10^{-2},\sigma_{mag}^2=1.000\times10^{-2}$. Additionally, the parameters $\alpha,\beta$, and $\kappa$, which determine the weights of the UKF, were set to $10^{-3}$, 2.0, and 0.0, respectively. The weights of the UKF are defined by \cite{b31}:
\begin{align}
    W_0^{(m)}=\frac{\lambda}{N_q+\lambda}\\
    W_0^{(c)}=\frac{\lambda}{N_q+\lambda}+(1-\alpha^2+\beta)\\
    W_i^{(m)}=W_i^{(c)}=\frac{1}{2(N_q+\lambda)}\quad (i\in\mathbb{Z},1\le i\le 2N_q)
\end{align}
$\lambda$ is calculated by $\lambda=\alpha^2(N_q+\kappa)-N_q$ \cite{b31}. In the KUKF, the magnetic reference vector was calculated using (\ref{eq:ref_mag}), and the covariance matrix of the process model was calculated using (\ref{eq:Qk}), which differs from \cite{b17}.
In the data preprocessing step, the acceleration data and magnetic data were low-pass filtered.

\begin{table*}[htbp]
  \begin{center}
    \caption{Performance of the MTW2-3A7G6 \cite{b20}}
    \small
      \begin{tabular}{|c|c|c|c|} \hline
          & Angular Velocity& Acceleration& Magnetic Field \\ \hline
        Axis& 3 axes& 3 axes& 3 axes\\ \hline
        Noise& $0.01deg/s/\sqrt{Hz}$& $200\mu g/\sqrt{Hz}$& $0.2mGauss/\sqrt{Hz}$\\ \hline
      \end{tabular}
    \label{table:Xsens_performance}
  \end{center}
\end{table*}

\begin{table*}[htbp]
  \begin{center}
    \caption{List of the collected data in \cite{b17}}
    \small
      \begin{tabular}{|c|c|c|c|c|} \hline
         Data Name & Walking Path& Walked Distance [m] & Measurement Time [s] \\ \hline
        Data 1-1& Straight-line walking& 4.62& 22.59\\ \hline
        Data 1-2& Straight-line walking& 4.60& 14.34\\ \hline
        Data 1-3& Straight-line walking& 4.54& 16.69\\ \hline
        Data 2-1& Walking along a rectangular path& 19.68& 30.3\\ \hline
        Data 2-2& Walking along a rectangular path& 19.90& 30.68\\ \hline
        Data 2-3& Walking along a rectangular path& 19.51& 28.37\\ \hline
        Data 3-1& Walking ten laps along a rectangular path& 200.70& 194.32\\ \hline
        Data 3-2& Walking ten laps along a rectangular path& 199.86& 200.2\\ \hline
        Data 3-3& Walking ten laps along a rectangular path& 202.89& 193.09\\ \hline
        Data 4-1& Walking one lap on the ground& 372.07& 366.27\\ \hline
        Data 4-2& Walking one lap on the ground& 372.89& 364.51\\ \hline
        Data 4-3& Walking one lap on the ground& 367.62& 327.61\\ \hline
        Data 5-1& Walking two laps on the ground& 757.38& 625.96\\ \hline
        Data 5-2& Walking two laps on the ground& 747.98& 611.15\\ \hline
        Data 5-3& Walking two laps on the ground& 763.42& 636.35\\ \hline
      \end{tabular}
    \label{table:data_details}
  \end{center}
\end{table*}

\subsection{Exp.1 Evaluating the Errors of the Kalman Filters}
\begin{figure}[htbp]
  \centering
  \includegraphics[width=90mm]{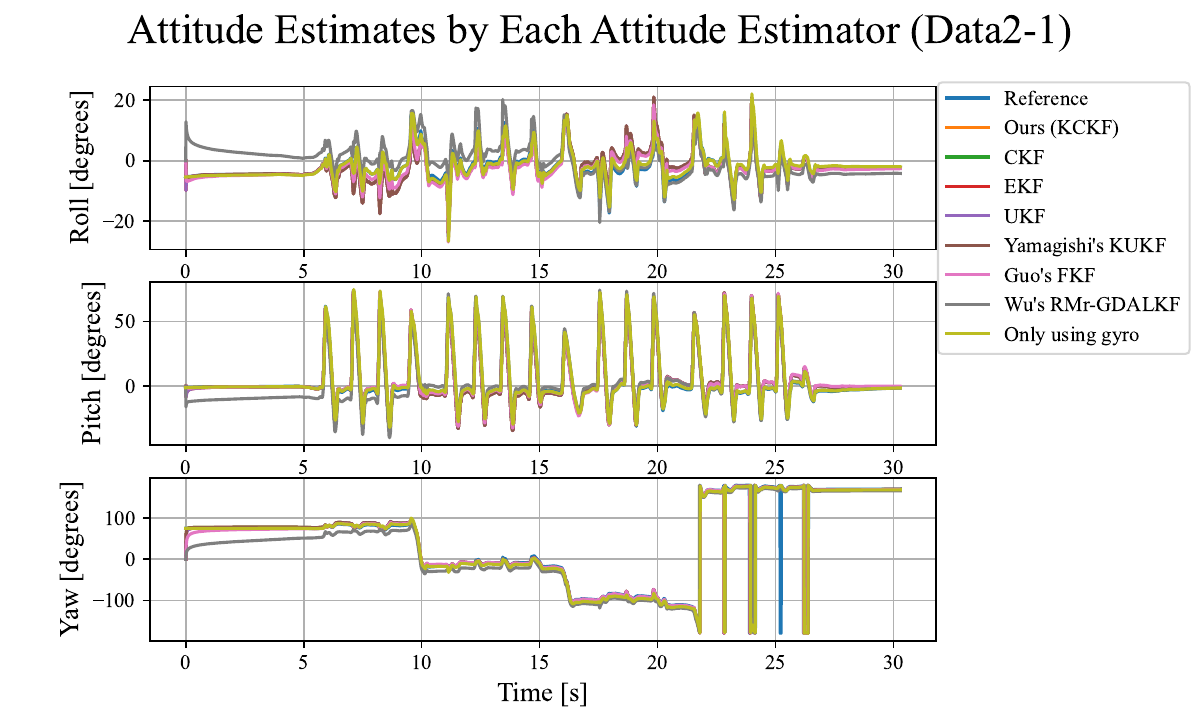}
  \caption{The estimation results of attitudes by each attitude estimator in Data 2-1}
  \label{fig:Attitude_Data2-1}
\end{figure}

\begin{figure}[htbp]
  \centering
  \includegraphics[width=90mm]{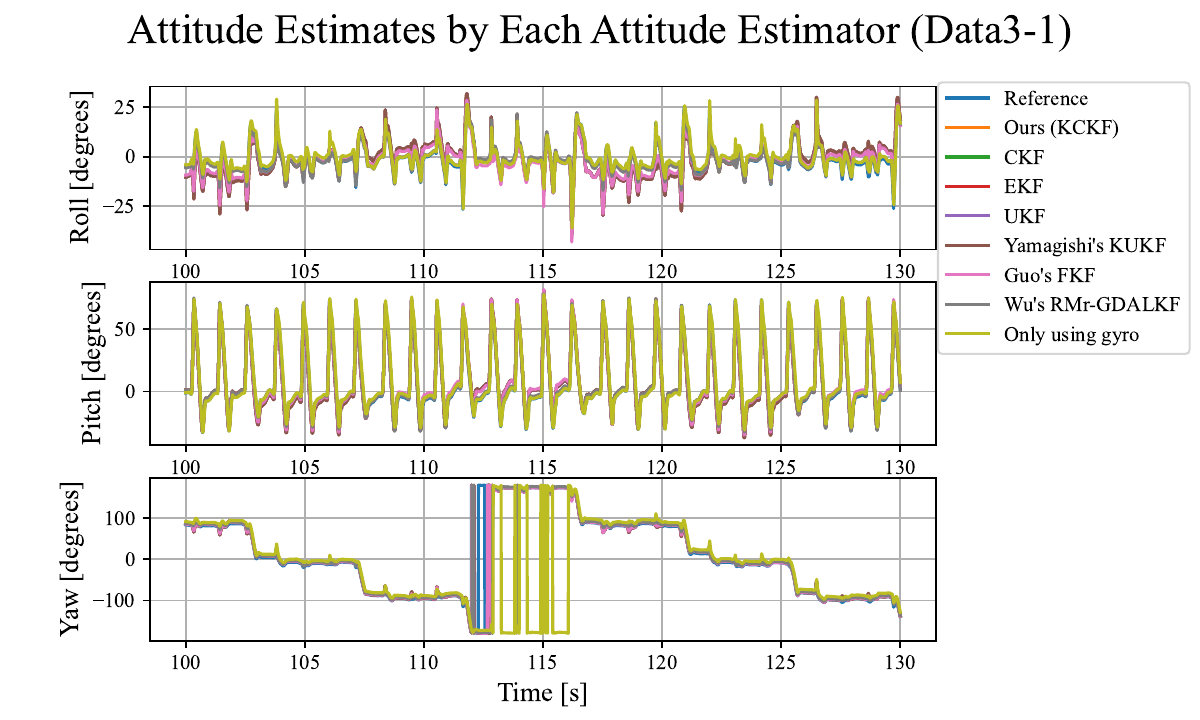}
  \caption{The estimation results of attitudes by each attitude estimator in Data 3-1 from 100 seconds to 130 seconds}
  \label{fig:Attitude_Data3-1}
\end{figure}

\begin{figure}[htbp]
  \centering
  \includegraphics[width=90mm]{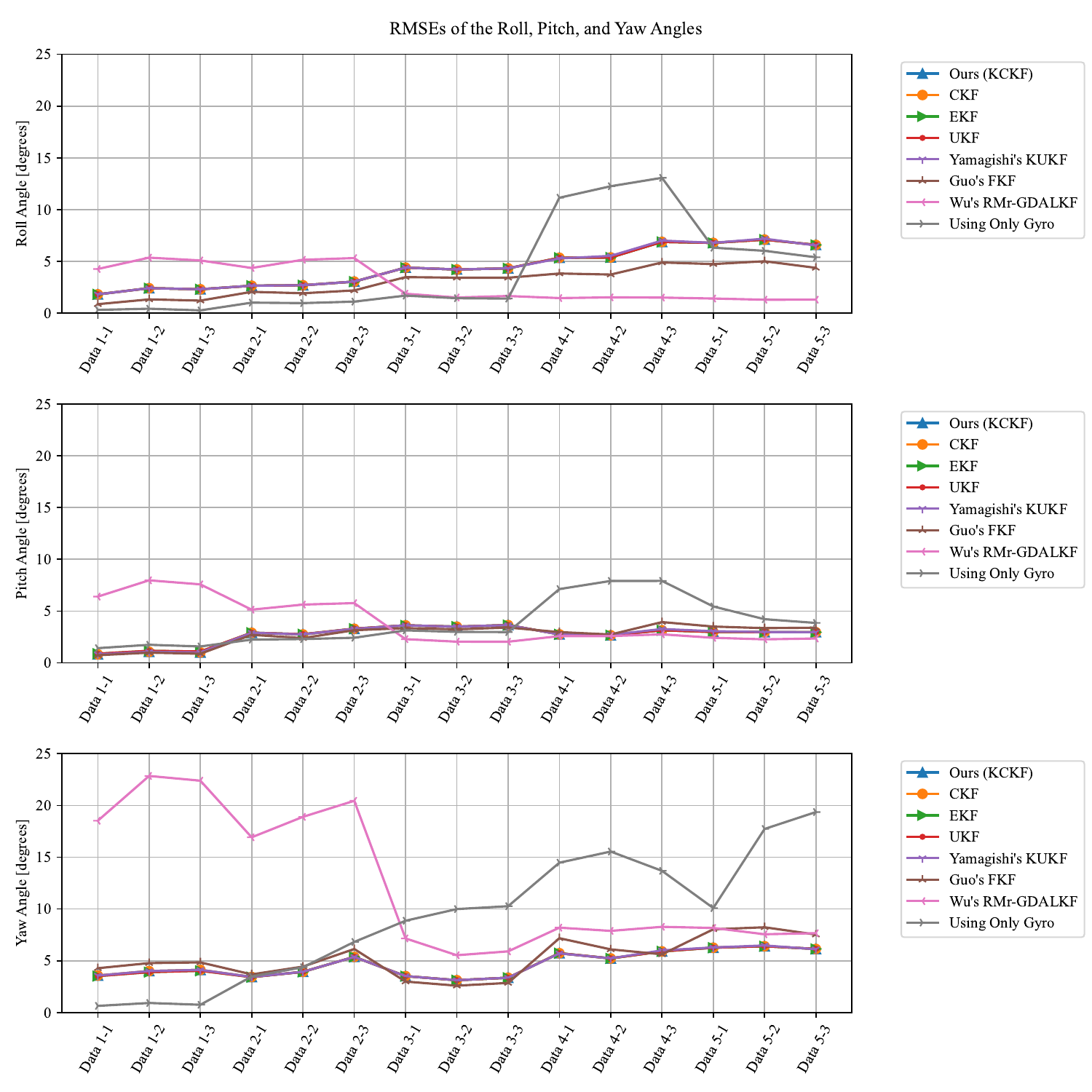}
  \caption{RMSEs of the Euler angles estimeted by each attitude estimator}
  \label{fig:RMSE_Eulers}
\end{figure}

Experiment 1 was conducted to evaluate the estimation errors of the Kalman filters and to compare the proposed KCKF with existing methods including CKF, EKF, UKF, KUKF (proposed by \cite{b17}), FKF (proposed by \cite{b1}), RMr-GDALKF (proposed by \cite{b13}). Figure \ref{fig:Attitude_Data2-1} and Figure \ref{fig:Attitude_Data3-1} show the estimation results by each attitude estimator. Figure \ref{fig:RMSE_Eulers} shows the RMSEs of the Euler angles estimated by each attitude estimator, and Table \ref{table:RMSE_of_Euler_Angles} shows the average RMSEs of the attitudes estimated by each Kalman filter from Data 1-1 to Data 5-3. The attitude of the AHRS estimated by the Xsens Kalman Filter (XKF3hm) \cite{b20} was used as the reference value, because XKF3hm provides high accuracy as an attitude estimator of the AHRS. Formula \ref{eq:RMSE} is used to evaluate the RMSEs of the Euler angles.
\begin{align}
    \label{eq:RMSE}
    \text{RMSE}=\sqrt{\frac{1}{n}\sum_{i=0}^ne_i^2}\\
    \text{, where }e_i^2=\min[{(\hat{y}_i-y_i^{ref})^2,\{\hat{y}_i-(y_i^{ref}\pm360^{\circ})\}^2}]
\end{align}
$n,\hat{y}_i$ denotes the length of the data, estimates by each attitude estimator. $y_i^{ref}$ denotes reference value estimated by XKF3hm \cite{b20}. From the experimental results, the KCKF maintains the accuracy of the CKF. Since the reference values are estimated by XKF3hm \cite{b20} and do not represent the complete ground truth, it is difficult to conclude which Kalman filter has the highest accuracy. However, according to the experimental results, the estimation errors of proposed KCKF are very close to those of CKF, EKF, UKF and KUKF.

\begin{table*}[htbp]
  \begin{center}
    \caption{The average RMSEs of the attitudes estimated by each Kalman filter from Data 1-1 to Data 5-3 (The values in parentheses are unbiased standard deviations; unit: degrees)}
    \small
      \begin{tabular}{|c|c|c|c|} \hline
          & Roll Angle& Pitch Angle& Yaw Angle \\ \hline
        Ours (KCKF) & 4.40 (1.87)& 2.67 (0.94)& 4.67 (1.21) \\ \hline
        CKF & 4.40 (1.87)& 2.67 (0.94)& 4.67 (1.21)\\ \hline
        EKF & 4.40 (1.86)& 2.70 (0.91)& 4.66 (1.23)\\ \hline
        UKF & 4.40 (1.86)& 2.70 (0.90)& 4.66 (1.23)\\ \hline
        KUKF & 4.42 (1.89)& 2.69 (0.94)& 4.69 (1.23)\\ \hline
        Guo's FKF & 3.10 (1.40)& 2.72 (1.03)& 5.29 (1.88)\\ \hline
        Wu's RMr-GDALKF & 2.88 (1.76)& 3.99 (2.17)& 12.43 (6.60)\\ \hline
        Using Only Gyro & 4.20 (4.61)& 3.82 (2.25)& 9.13 (6.22)\\ \hline
      \end{tabular}
    \label{table:RMSE_of_Euler_Angles}
  \end{center}
\end{table*}

\subsection{Exp.2 Comparing the Computation Time of the Kalman Filters}
In Experiment 2, using Data 5-1, the computation time of KCKF was compared with those of the CKF, EKF, UKF, KUKF (proposed by \cite{b17}), FKF (proposed by \cite{b1}), and RMr-GDALKF (proposed by \cite{b13}). Experiment 2 was carried out in both the laptop environment (MacBook Pro 2021, M1 Pro chip, 14-inch \cite{b38}) and the low-cost computer environment (Raspberry Pi 4 Model B \cite{b39}).\par
\begin{figure}[htbp]
  \centering
  \includegraphics[width=78mm]{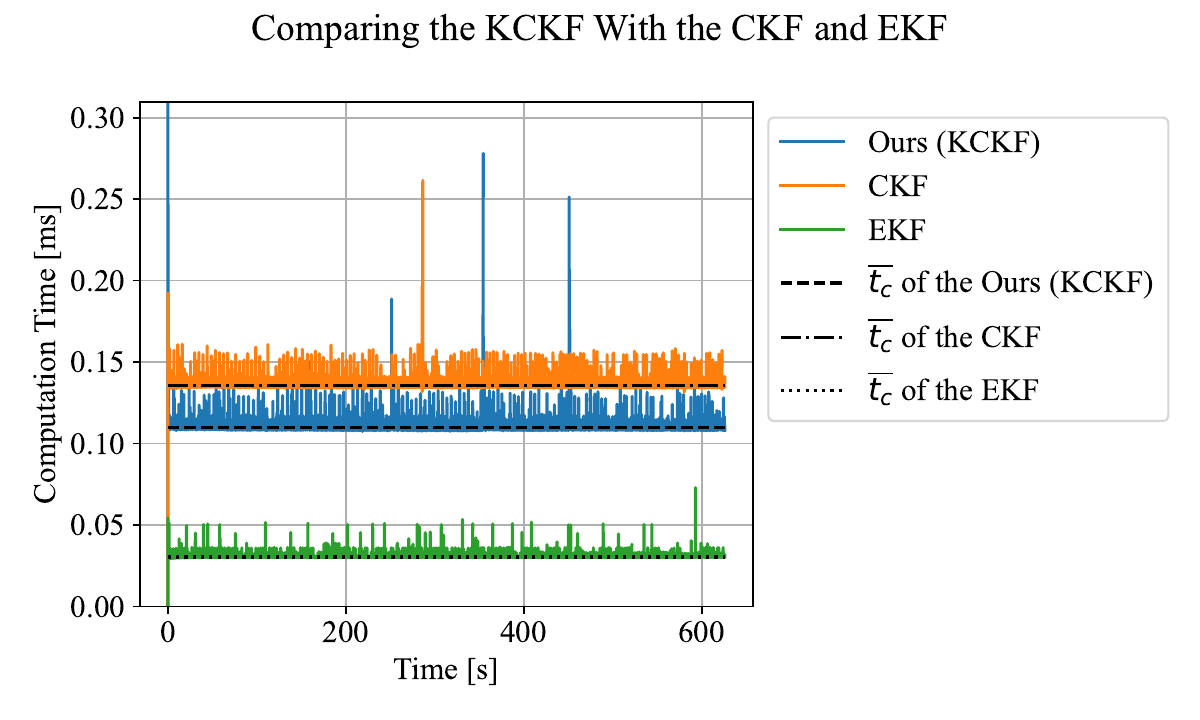}
  \caption{Comparing the KCKF with the CKF and EKF (MacBook Pro 2021 environment)}
  \label{fig:computation_time1_mac}
\end{figure}
\begin{figure}[htbp]
  \centering
  \includegraphics[width=78mm]{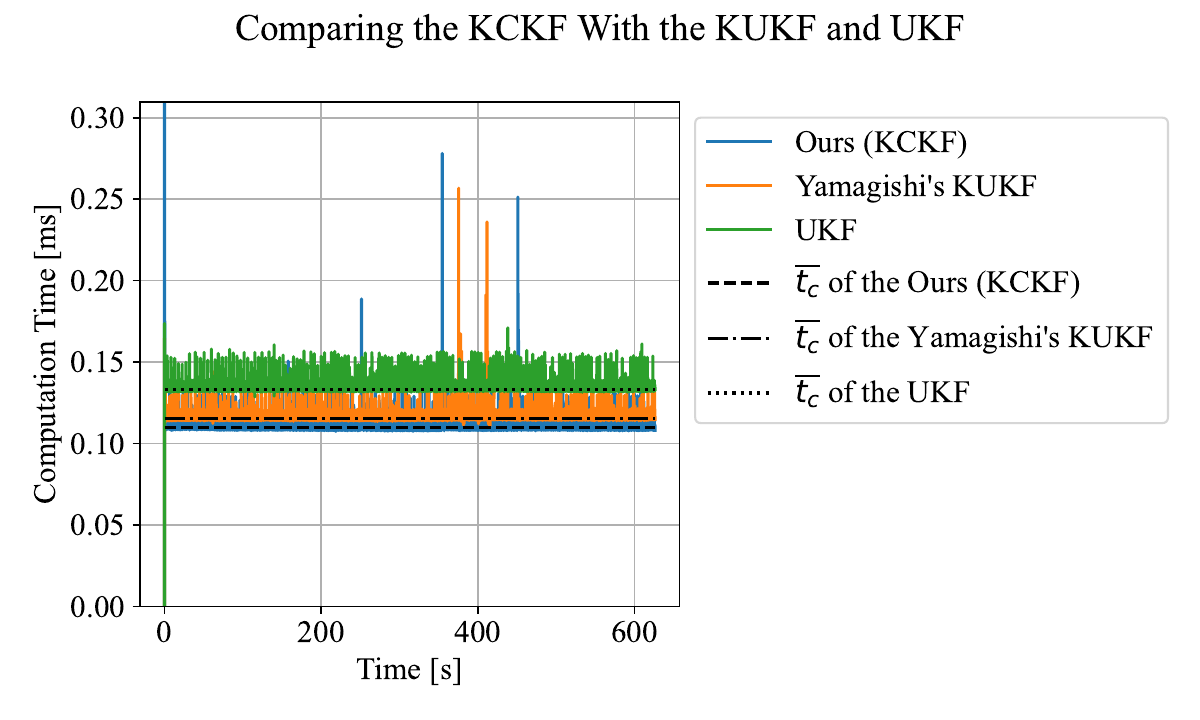}
  \caption{Comparing the KCKF with the KUKF and UKF (MacBook Pro 2021 environment)}
  \label{fig:computation_time2_mac}
\end{figure}
\begin{figure}[htbp]
  \centering
  \includegraphics[width=78mm]{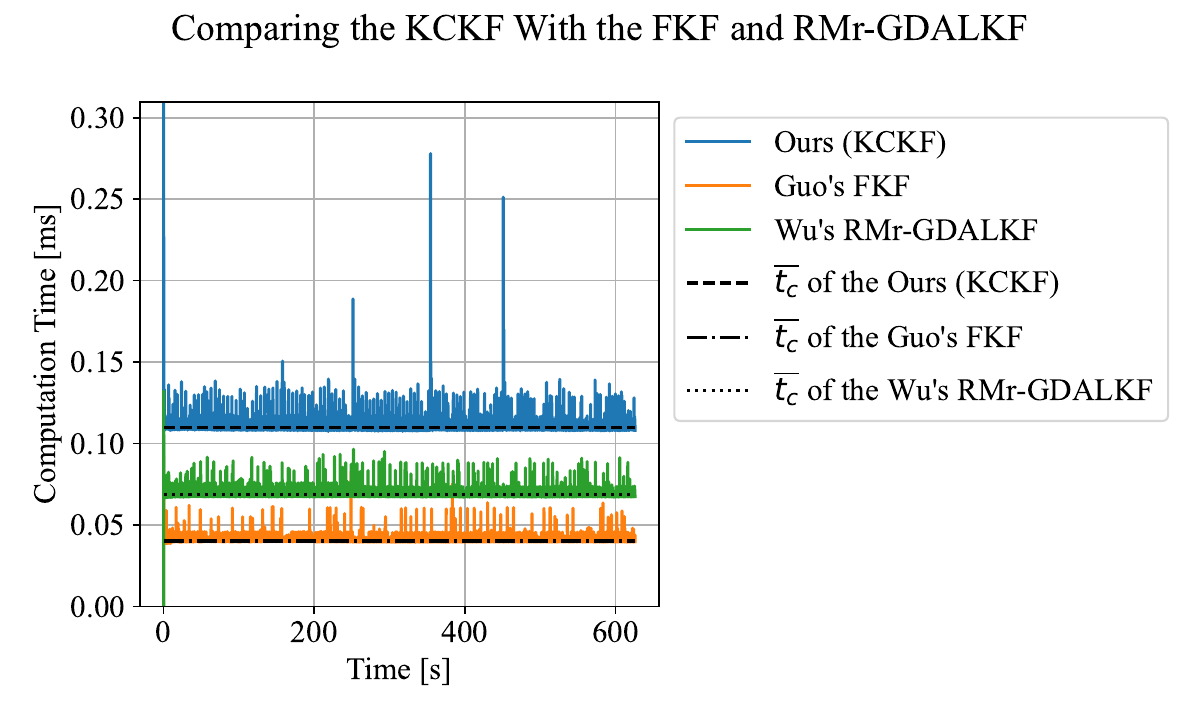}
  \caption{Comparing the KCKF with the Guo's FKF and Wu's RMr-GDALKF (MacBook Pro 2021 environment)}
  \label{fig:computation_time3_mac}
\end{figure}
\begin{figure}[htbp]
  \centering
  \includegraphics[width=78mm]{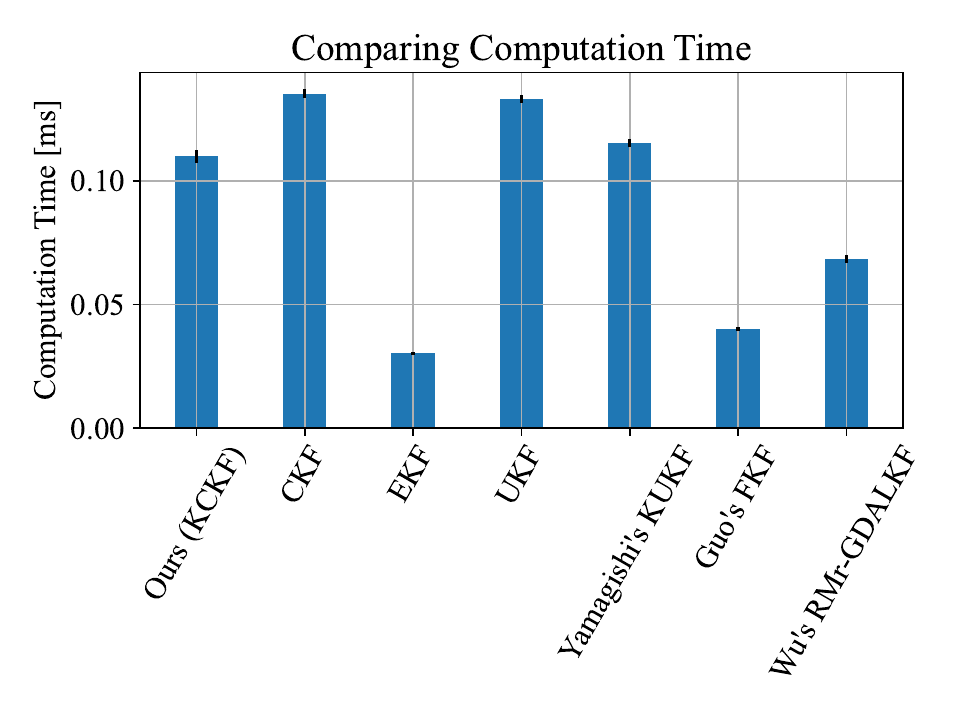}
  \caption{A bar chart of the average computation time $\overline{t_c}$ for each Kalman filter (MacBook Pro 2021 environment)}
  \label{fig:computation_time_bar_mac}
\end{figure}

Figures \ref{fig:computation_time1_mac}, \ref{fig:computation_time2_mac} and \ref{fig:computation_time3_mac} show the computation time of each Kalman filter in the MacBook environment. Figure \ref{fig:computation_time_bar_mac} shows a bar chart summarizing the data shown in Figures \ref{fig:computation_time1_mac}, \ref{fig:computation_time2_mac} and \ref{fig:computation_time3_mac}. Table \ref{table:computation_time_mac} shows the numerical results of computation time measurements in the MacBook Pro environment.$\overline{t_c}$ denotes the average computation time for each Kalman filter to process a measurement. In this paper, we count a set of nine-axis data including acceleration, angular velocity, and magnetometer data as a measurement.\par
\begin{figure}[htbp]
  \centering
  \includegraphics[width=78mm]{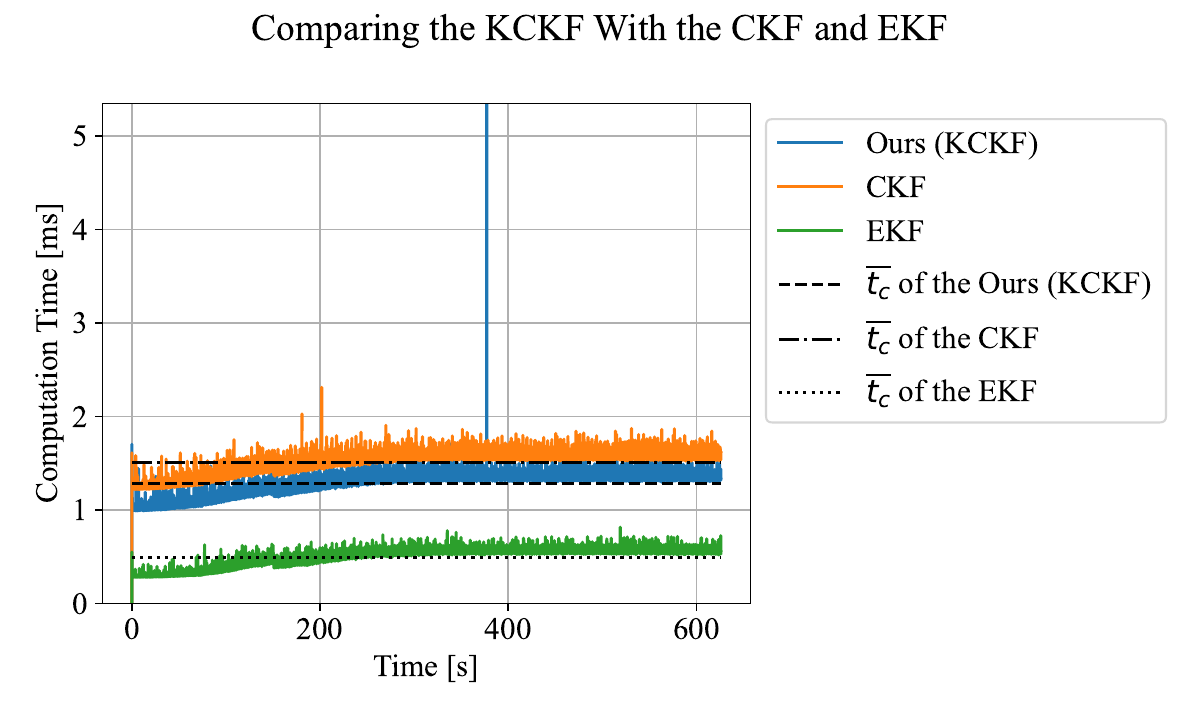}
  \caption{Comparing the KCKF with the CKF and EKF (Raspberry Pi 4 environment)}
  \label{fig:computation_time1_RasPi}
\end{figure}
\begin{figure}[htbp]
  \centering
  \includegraphics[width=78mm]{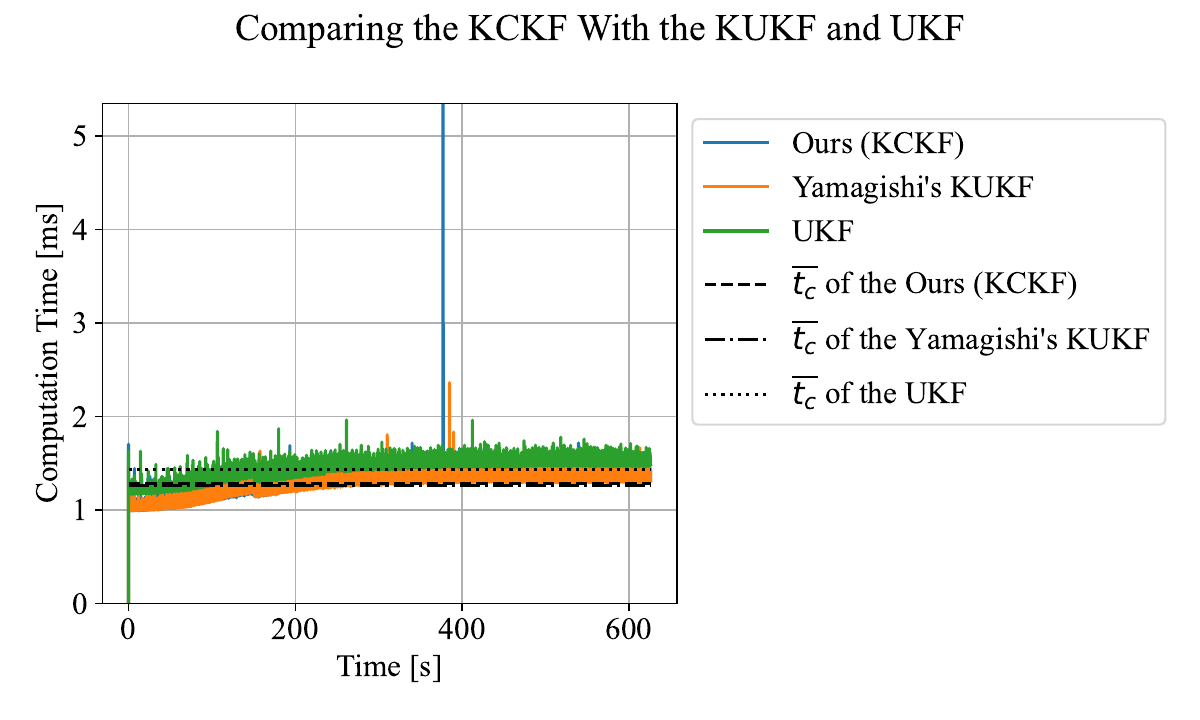}
  \caption{Comparing the KCKF with the KUKF and UKF (Raspberry Pi 4 environment)}
  \label{fig:computation_time2_RasPi}
\end{figure}
\begin{figure}[htbp]
  \centering
  \includegraphics[width=78mm]{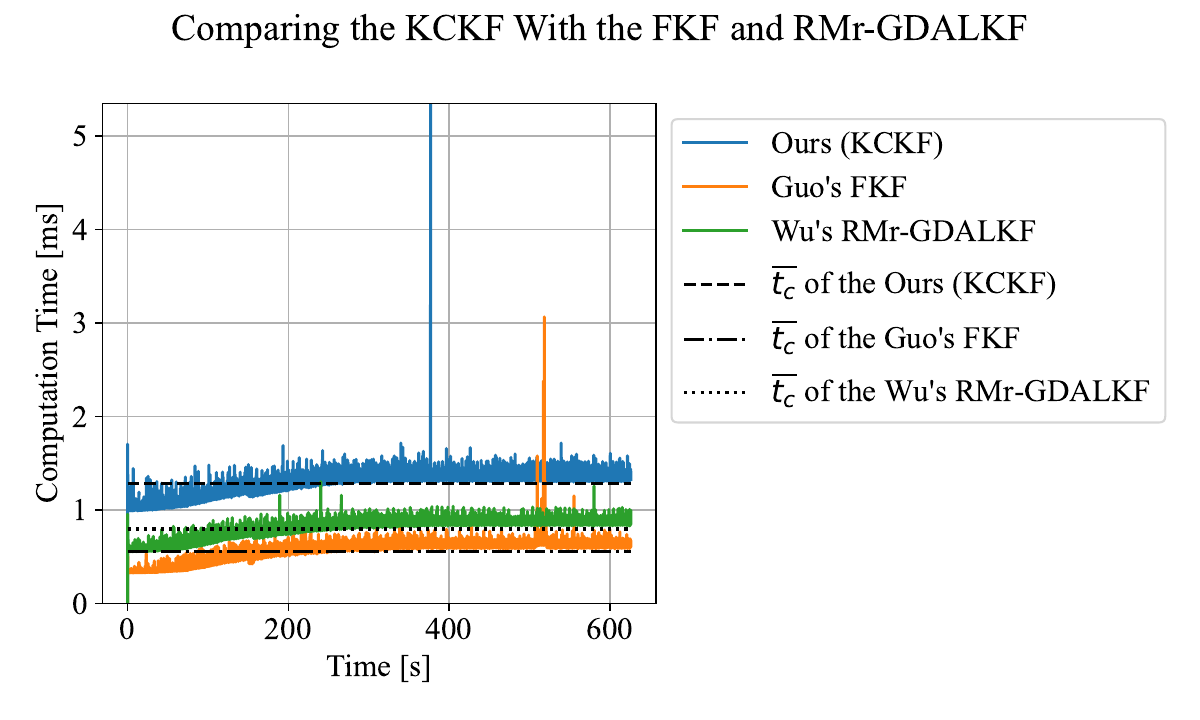}
  \caption{Comparing the KCKF with the Guo's FKF and Wu's RMr-GDALKF (Raspberry Pi 4 environment)}
  \label{fig:computation_time3_RasPi}
\end{figure}
\begin{figure}[htbp]
  \centering
  \includegraphics[width=78mm]{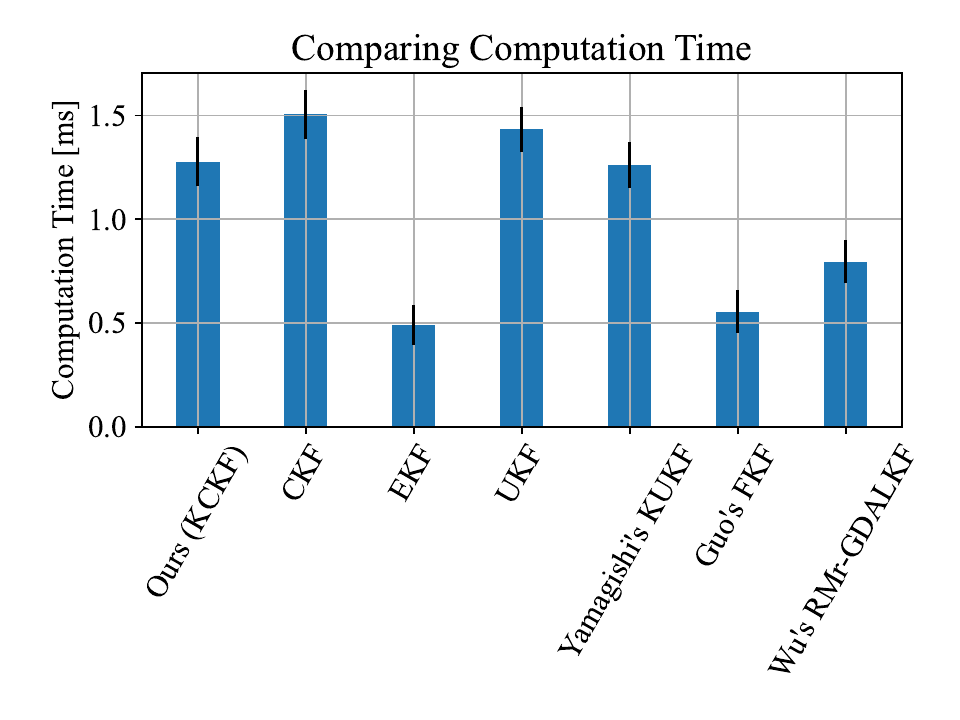}
  \caption{A bar chart of the average computation time $\overline{t_c}$ for each Kalman filter (Raspberry Pi 4 environment)}
  \label{fig:computation_time_bar_RasPi}
\end{figure}

Figures \ref{fig:computation_time1_RasPi}, \ref{fig:computation_time2_RasPi} and \ref{fig:computation_time3_RasPi} show the computation time of each Kalman filter in the Raspberry Pi 4 Model B environment. Figure \ref{fig:computation_time_bar_RasPi} shows a bar chart summarizing the data shown in Figures \ref{fig:computation_time1_RasPi}, \ref{fig:computation_time2_RasPi} and \ref{fig:computation_time3_RasPi}. Table \ref{table:computation_time_RasPi} shows the numerical results of computation time measurements in the Raspberry Pi 4 Model B environment.\par
From the experimental results, the KCKF reduced the computation time by approximately 18.79\% compared to the CKF in the MacBook Pro environment, whereas the KCKF reduced the computation time by approximately 15.15\% compared to the CKF in the Raspberry Pi 4 Model B environment. In addition, the KCKF was slightly faster than the KUKF in the MacBook environment, but the computation time of the KCKF and that of the KUKF were almost the same in Raspberry Pi 4 environment. However, the KCKF required substantially more computation time compared to the EKF, FKF \cite{b1}, and RMr-GDALKF \cite{b13}. These experimental results show that the EKF required the lowest computational cost.

\begin{table}[htbp]
  \begin{center}
    \caption{Numerical results of computation time measurements in the MacBook Pro environment (USD: unbiased standard deviation; unit: ms)}
    \small
      \begin{tabular}{|c|c|c|} \hline
          & $\overline{t_c}$&  USD\\ \hline
        Ours (KCKF)& $1.10\times10^{-1}$& $2.63\times10^{-3}$\\ \hline
        CKF&  $1.35\times10^{-1}$& $1.67\times10^{-3}$\\ \hline
        EKF& $3.03\times10^{-2}$& $6.84\times10^{-4}$\\ \hline
        UKF& $1.33\times10^{-1}$& $1.61\times10^{-3}$\\ \hline
        Yamagishi's KUKF& $1.15\times10^{-1}$& $1.78\times10^{-3}$\\ \hline
        Guo's FKF& $4.01\times10^{-2}$ & $9.57\times10^{-4}$\\ \hline
        Wu's RMr-GDALKF& $6.84\times10^{-2}$ & $1.51\times10^{-3}$\\ \hline
      \end{tabular}
    \label{table:computation_time_mac}
  \end{center}
\end{table}

\begin{table}[htbp]
  \begin{center}
    \caption{Numerical results of computation time measurements in the Raspberry Pi 4 environment (USD: unbiased standard deviation; unit: ms)}
    \small
      \begin{tabular}{|c|c|c|} \hline
          & $\overline{t_c}$&  USD\\ \hline
        Ours (KCKF)& $1.28$& $1.19\times10^{-1}$\\ \hline
        CKF&  $1.51$& $1.17\times10^{-1}$\\ \hline
        EKF& $4.92\times10^{-1}$& $9.61\times10^{-2}$\\ \hline
        UKF& $1.43$& $1.09\times10^{-1}$\\ \hline
        Yamagishi's KUKF& $1.26$& $1.12\times10^{-1}$\\ \hline
        Guo's FKF& $5.55\times10^{-1}$ & $1.03\times10^{-1}$\\ \hline
        Wu's RMr-GDALKF& $7.95\times10^{-1}$ & $1.04\times10^{-1}$\\ \hline
      \end{tabular}
    \label{table:computation_time_RasPi}
  \end{center}
\end{table}

\section{Conclusion}
In this paper, the KCKF, which has a lower computational cost than the CKF, was proposed, and the KCKF was compared with the CKF, UKF, KUKF, and EKF. The experimental results showed that the KCKF reduced the computation time by approximately 18.79\% compared to the CKF in the MacBook Pro 2021 environment, whereas the KCKF reduced the computation time by 15.15\% compared to the CKF in the Raspberry Pi 4 Model B environment. In addition, the KCKF maintained the accuracy of the CKF.\par
However, this paper has limitations. First, since the reference values used to evaluate the RMSEs of the Euler angles were the attitudes estimated by XKF3hm \cite{b20} and do not represent complete ground truth, it is difficult to conclude which Kalman filter has the highest accuracy. Second, we should verify the accuracy of the CKF under other conditions, such as more complex movements, and compare it with that of the EKF. According to our experimental results, the accuracies of the KCKF and CKF are very close to those of the EKF, and it cannot be concluded from the experimental results that the KCKF and CKF are more accurate than the EKF. However, in theory, the CKF has higher accuracy than the EKF. Additionally, KCKF and CKF have an advantage of being Jacobian-free. Third, in the experiments of this paper, only the pedestrian walking data were used, and we do not know whether the KCKF can maintain the accuracy of the CKF under conditions such as higher external accelerations and higher magnetic noise.\par
In future work, we should compare the KCKF with representative Kalman filters using a more accurate reference value as ground truth. We also need to compare the KCKF with representative Kalman filters under the conditions of higher external accelerations, higher magnetic noise, and more complex movements. For example, we should test the accuracy of the KCKF using indoor walking and running data. We should further develop the KCKF to reduce computational cost and improve accuracy.

\FloatBarrier
\appendices
\allowdisplaybreaks
\section{Proof}
We prove that the formulas from (\ref{eq:KCKF1}) to (\ref{eq:KCKF4}) provide the same computational results as the formulas from (\ref{eq:CKF1}) to (\ref{eq:CKF4}). The formulas from (\ref{eq:KCKF1}) to (\ref{eq:KCKF4}) are derived from the formulas from (\ref{eq:CKF1}) to (\ref{eq:CKF4}).\par
First, we prove that (\ref{eq:KCKF2}) is derived from (\ref{eq:CKF3}).
    \begin{align}
        &\hat{\boldsymbol{q}}_{k|k-1}=\frac{1}{2N_q}\sum_{i=1}^{2N_q}\boldsymbol{\mathcal{Q}}^*_{k|k-1}\\
        &=\frac{1}{2N_q}\sum_{i=1}^{2N_q}\boldsymbol{f}(\boldsymbol{\mathcal{Q}}^{(i)}_{k-1|k-1})\\
        &=\frac{1}{2N_q}\sum_{i=1}^{2N_q}\boldsymbol{F}_k(\boldsymbol{S}_{k-1|k-1}\boldsymbol{e}_i+\hat{\boldsymbol{q}}_{k-1|k-1})\\
        &=\frac{1}{2N_q}\sum_{i=1}^{2N_q}(\boldsymbol{F}_k\boldsymbol{S}_{k-1|k-1}\boldsymbol{e}_i+\boldsymbol{F}_k\hat{\boldsymbol{q}}_{k-1|k-1})\\
        &=\boldsymbol{F}_k\hat{\boldsymbol{q}}_{k-1|k-1}+\frac{1}{2N_q}\sum_{i=1}^{2N_q}\boldsymbol{F}_k\boldsymbol{S}_{k-1|k-1}\boldsymbol{e}_i\\
        &=\boldsymbol{F}_k\hat{\boldsymbol{q}}_{k-1|k-1}+\frac{1}{2N_q}\Bigg(\sum_{i=1}^{N_q}\boldsymbol{F}_k\boldsymbol{S}_{k-1|k-1}\boldsymbol{e}_i\\
        &\quad+\sum_{i=N_q+1}^{2N_q}\boldsymbol{F}_k\boldsymbol{S}_{k-1|k-1}\boldsymbol{e}_i\Bigg) \notag \\
        &=\boldsymbol{F}_k\hat{\boldsymbol{q}}_{k-1|k-1}+\frac{1}{2N_q}\Bigg(\sum_{i=1}^{N_q}\boldsymbol{F}_k\boldsymbol{S}_{k-1|k-1}\boldsymbol{e}_i\\
        &\quad-\sum_{i=1}^{N_q}\boldsymbol{F}_k\boldsymbol{S}_{k-1|k-1}\boldsymbol{e}_i\Bigg) \notag \\
        &=\boldsymbol{F}\hat{\boldsymbol{q}}_{k-1|k-1}
    \end{align}
    Therefore, (\ref{eq:KCKF2}) is derived from (\ref{eq:CKF3}), and (\ref{eq:KCKF2}) and (\ref{eq:CKF3}) have an equivalence relation.\par
    Then, we prove that (\ref{eq:KCKF3}) is derived from (\ref{eq:CKF2}).
    \begin{align}
        \boldsymbol{\mathcal{Q}}^{(i)}_{k|k-1}&=\boldsymbol{f}(\boldsymbol{\mathcal{Q}}^{(i)}_{k-1|k-1})\\
        &=\boldsymbol{F}_k(\boldsymbol{S}_{k-1|k-1}\boldsymbol{e}_i+\hat{\boldsymbol{q}}_{k-1|k-1})\\
        &=\boldsymbol{F}_k\boldsymbol{S}_{k-1|k-1}\boldsymbol{e}_i+\boldsymbol{F}_k\hat{\boldsymbol{q}}_{k-1|k-1}\\
    &\text{Here, let }\boldsymbol{M}_k=\boldsymbol{F}_k\boldsymbol{S}_{k-1|k-1}\text{, then,}\\
        \boldsymbol{\mathcal{Q}}^{(i)}_{k|k-1}&=\boldsymbol{M}_k\boldsymbol{e}_i+\hat{\boldsymbol{q}}_{k|k-1}
    \end{align}
    Therefore, (\ref{eq:KCKF3}) is derived from (\ref{eq:CKF2}), and (\ref{eq:KCKF3}) and (\ref{eq:CKF2}) have an equivalence relation.\par
    Finally, we prove that (\ref{eq:KCKF4}) is derived from (\ref{eq:CKF4}).
    \begin{align}
        \boldsymbol{P}_{k|k-1}&=\frac{1}{2N_q}\sum_{i=1}^{2N_q}\boldsymbol{\mathcal{Q}}^{*(i)}_{k|k-1}{\boldsymbol{\mathcal{Q}}^{*(i)}_{k|k-1}}^T-\hat{\boldsymbol{q}}_{k|k-1}\hat{\boldsymbol{q}}_{k|k-1}^T+\boldsymbol{Q}_k
    \end{align}
    Here, let $\boldsymbol{\mathcal{Q}}^s_k=\frac{1}{2N_q}\sum_{i=1}^{2N_q}\boldsymbol{\mathcal{Q}}^{*(i)}_{k|k-1}{\boldsymbol{\mathcal{Q}}^{*(i)}_{k|k-1}}^T$.
    \begin{align}
        &\boldsymbol{\mathcal{Q}}^s_k=\frac{1}{2N_q}\sum_{i=1}^{2N_q}\boldsymbol{\mathcal{Q}}^{*(i)}_{k|k-1}{\boldsymbol{\mathcal{Q}}^{*(i)}_{k|k-1}}^T\\
        &=\frac{1}{2N_q}\sum_{i=1}^{2N_q}\boldsymbol{f}(\boldsymbol{\mathcal{Q}}^{(i)}_{k-1|k-1})\{\boldsymbol{f}(\boldsymbol{\mathcal{Q}}^{(i)}_{k-1|k-1})\}^T\\
        &=\frac{1}{2N_q}\sum_{i=1}^{2N_q}\boldsymbol{F}_k(\boldsymbol{S}_{k-1|k-1}\boldsymbol{e}_i+\hat{\boldsymbol{q}}_{k-1|k-1})\\
        &\quad\times\{\boldsymbol{F}_k(\boldsymbol{S}_{k-1|k-1}\boldsymbol{e}_i+\hat{\boldsymbol{q}}_{k-1|k-1})\}^T \notag \\
        &=\frac{1}{2N_q}\sum_{i=1}^{2N_q}\boldsymbol{F}_k(\boldsymbol{S}_{k-1|k-1}\boldsymbol{e}_i+\hat{\boldsymbol{q}}_{k-1|k-1})\\
        &\quad\times(\boldsymbol{S}_{k-1|k-1}\boldsymbol{e}_i+\hat{\boldsymbol{q}}_{k-1|k-1})^T\boldsymbol{F}_k^T \notag \\
        &=\frac{1}{2N_q}\sum_{i=1}^{2N_q}\boldsymbol{F}_k(\boldsymbol{S}_{k-1|k-1}\boldsymbol{e}_i+\hat{\boldsymbol{q}}_{k-1|k-1})\\
        &\quad\times(\boldsymbol{e}_i^T\boldsymbol{S}_{k-1|k-1}^T+\hat{\boldsymbol{q}}_{k-1|k-1}^T)\boldsymbol{F}_k^T \notag \\
        &=\frac{1}{2N_q}\sum_{i=1}^{2N_q}\boldsymbol{F}_k(\boldsymbol{S}_{k-1|k-1}\boldsymbol{e}_i\boldsymbol{e}_i^T\boldsymbol{S}_{k-1|k-1}^T\\
        &\quad+\hat{\boldsymbol{q}}_{k-1|k-1}\boldsymbol{e}_i^T\boldsymbol{S}_{k-1|k-1}^T+\boldsymbol{S}_{k-1|k-1}\boldsymbol{e}_i\hat{\boldsymbol{q}}_{k-1|k-1}^T \notag \\
        &\quad+\hat{\boldsymbol{q}}_{k-1|k-1}\hat{\boldsymbol{q}}_{k-1|k-1}^T)\boldsymbol{F}_k^T \notag \\
        &=\frac{1}{2N_q}\sum_{i=1}^{N_q}\boldsymbol{F}_k(\boldsymbol{S}_{k-1|k-1}\boldsymbol{e}_i\boldsymbol{e}_i^T\boldsymbol{S}_{k-1|k-1}^T\\
        &\quad+\hat{\boldsymbol{q}}_{k-1|k-1}\boldsymbol{e}_i^T\boldsymbol{S}_{k-1|k-1}^T \notag \\
        &\quad+\boldsymbol{S}_{k-1|k-1}\boldsymbol{e}_i\hat{\boldsymbol{q}}_{k-1|k-1}^T+\hat{\boldsymbol{q}}_{k-1|k-1}\hat{\boldsymbol{q}}_{k-1|k-1}^T)\boldsymbol{F}_k^T \notag \\
        &\quad+\frac{1}{2N_q}\sum_{i=N_q+1}^{2N_q}\boldsymbol{F}_k(\boldsymbol{S}_{k-1|k-1}\boldsymbol{e}_i\boldsymbol{e}_i^T\boldsymbol{S}_{k-1|k-1}^T \notag \\
        &\quad+\hat{\boldsymbol{q}}_{k-1|k-1}\boldsymbol{e}_i^T\boldsymbol{S}_{k-1|k-1}^T+\boldsymbol{S}_{k-1|k-1}\boldsymbol{e}_i\hat{\boldsymbol{q}}_{k-1|k-1}^T \notag \\
        &\quad+\hat{\boldsymbol{q}}_{k-1|k-1}\hat{\boldsymbol{q}}_{k-1|k-1}^T)\boldsymbol{F}_k^T \notag \\
        &=\frac{1}{2N_q}\sum_{i=1}^{N_q}\boldsymbol{F}_k(\boldsymbol{S}_{k-1|k-1}\boldsymbol{e}_i\boldsymbol{e}_i^T\boldsymbol{S}_{k-1|k-1}^T\\
        &\quad+\hat{\boldsymbol{q}}_{k-1|k-1}\boldsymbol{e}_i^T\boldsymbol{S}_{k-1|k-1}^T+\boldsymbol{S}_{k-1|k-1}\boldsymbol{e}_i\hat{\boldsymbol{q}}_{k-1|k-1}^T \notag \\
        &\quad+\hat{\boldsymbol{q}}_{k-1|k-1}\hat{\boldsymbol{q}}_{k-1|k-1}^T)\boldsymbol{F}_k^T \notag \\
        &\quad+\frac{1}{2N_q}\sum_{i=1}^{N_q}\boldsymbol{F}_k(\boldsymbol{S}_{k-1|k-1}\boldsymbol{e}_i\boldsymbol{e}_i^T\boldsymbol{S}_{k-1|k-1}^T \notag \\
        &\quad-\hat{\boldsymbol{q}}_{k-1|k-1}\boldsymbol{e}_i^T\boldsymbol{S}_{k-1|k-1}^T-\boldsymbol{S}_{k-1|k-1}\boldsymbol{e}_i\hat{\boldsymbol{q}}_{k-1|k-1}^T \notag \\
        &\quad+\hat{\boldsymbol{q}}_{k-1|k-1}\hat{\boldsymbol{q}}_{k-1|k-1}^T)\boldsymbol{F}_k^T \notag \\
        &=2\times\frac{1}{2N_q}\sum_{i=1}^{N_q}\boldsymbol{F}_k(\boldsymbol{S}_{k-1|k-1}\boldsymbol{e}_i\boldsymbol{e}_i^T\boldsymbol{S}_{k-1|k-1}^T\\
        &\quad+\hat{\boldsymbol{q}}_{k-1|k-1}\hat{\boldsymbol{q}}_{k-1|k-1}^T)\boldsymbol{F}_k^T \notag \\
        &=\frac{1}{N_q}\biggl\{\boldsymbol{F}_k\boldsymbol{S}_{k-1|k-1}\biggr(\sum_{i=1}^{N_q}\boldsymbol{e}_i\boldsymbol{e}_i^T\biggl)\boldsymbol{S}_{k-1|k-1}^T\boldsymbol{F}_k^T\\
        &\quad+\boldsymbol{F}_k\biggl(\sum_{i=1}^{N_q}\hat{\boldsymbol{q}}_{k-1|k-1}\hat{\boldsymbol{q}}_{k-1|k-1}^T\biggr)\boldsymbol{F}_k^T\biggr\} \notag \\
        &=\frac{1}{N_q}\{\boldsymbol{F}_k\boldsymbol{S}_{k-1|k-1}(N_q\boldsymbol{I}_{4\times4})\boldsymbol{S}_{k-1|k-1}^T\boldsymbol{F}_k^T\\
        &\quad+\boldsymbol{F}_k(N_q\hat{\boldsymbol{q}}_{k-1|k-1}\hat{\boldsymbol{q}}_{k-1|k-1}^T)\boldsymbol{F}_k^T\} \notag \\
        &=\boldsymbol{F}_k\boldsymbol{S}_{k-1|k-1}(\boldsymbol{F}_k\boldsymbol{S}_{k-1|k-1})^T+\boldsymbol{F}_k\hat{\boldsymbol{q}}_{k-1|k-1}\hat{\boldsymbol{q}}_{k-1|k-1}^T\boldsymbol{F}_k^T\\
        &=\boldsymbol{M}_k\boldsymbol{M}_k^T+\boldsymbol{F}_k\hat{\boldsymbol{q}}_{k-1|k-1}\hat{\boldsymbol{q}}_{k-1|k-1}^T\boldsymbol{F}_k^T\\
        &=\boldsymbol{M}_k\boldsymbol{M}_k^T+\boldsymbol{F}_k\hat{\boldsymbol{q}}_{k-1|k-1}(\boldsymbol{F}_k\hat{\boldsymbol{q}}_{k-1|k-1})^T\\
        &=\boldsymbol{M}_k\boldsymbol{M}_k^T+\hat{\boldsymbol{q}}_{k|k-1}\hat{\boldsymbol{q}}_{k|k-1}^T
    \end{align}
    Therefore,
    \begin{align}
        \boldsymbol{P}_{k|k-1}&=\frac{1}{2N_q}\sum_{i=1}^{2N_q}\boldsymbol{\mathcal{Q}}^{*(i)}_{k|k-1}{\boldsymbol{\mathcal{Q}}^{*(i)}_{k|k-1}}^T-\hat{\boldsymbol{q}}_{k|k-1}\hat{\boldsymbol{q}}_{k|k-1}^T+\boldsymbol{Q}_k\\
        &=\boldsymbol{\mathcal{Q}}^s_k-\hat{\boldsymbol{q}}_{k|k-1}\hat{\boldsymbol{q}}_{k|k-1}^T+\boldsymbol{Q}_k\\
        &=(\boldsymbol{M}_k\boldsymbol{M}_k^T+\hat{\boldsymbol{q}}_{k|k-1}\hat{\boldsymbol{q}}_{k|k-1}^T)-\hat{\boldsymbol{q}}_{k|k-1}\hat{\boldsymbol{q}}_{k|k-1}^T+\boldsymbol{Q}_k\\
        &=\boldsymbol{M}_k\boldsymbol{M}_k^T+\boldsymbol{Q}_k
    \end{align}
    Thus, (\ref{eq:KCKF3}) is derived from (\ref{eq:CKF2}), and (\ref{eq:KCKF3}) and (\ref{eq:CKF2}) have an equivalence relation.\par
    Therefore, the formulas from (\ref{eq:KCKF1}) to (\ref{eq:KCKF4}) provide the same computational results as the formulas from (\ref{eq:CKF1}) to (\ref{eq:CKF4}), because they have equivalence relation. $\Box$

\nocite{*}
\bibliography{refs}
\bibliographystyle{IEEEtran}

\begin{IEEEbiography}[{\includegraphics[width=1in,height=1.25in,clip,keepaspectratio]{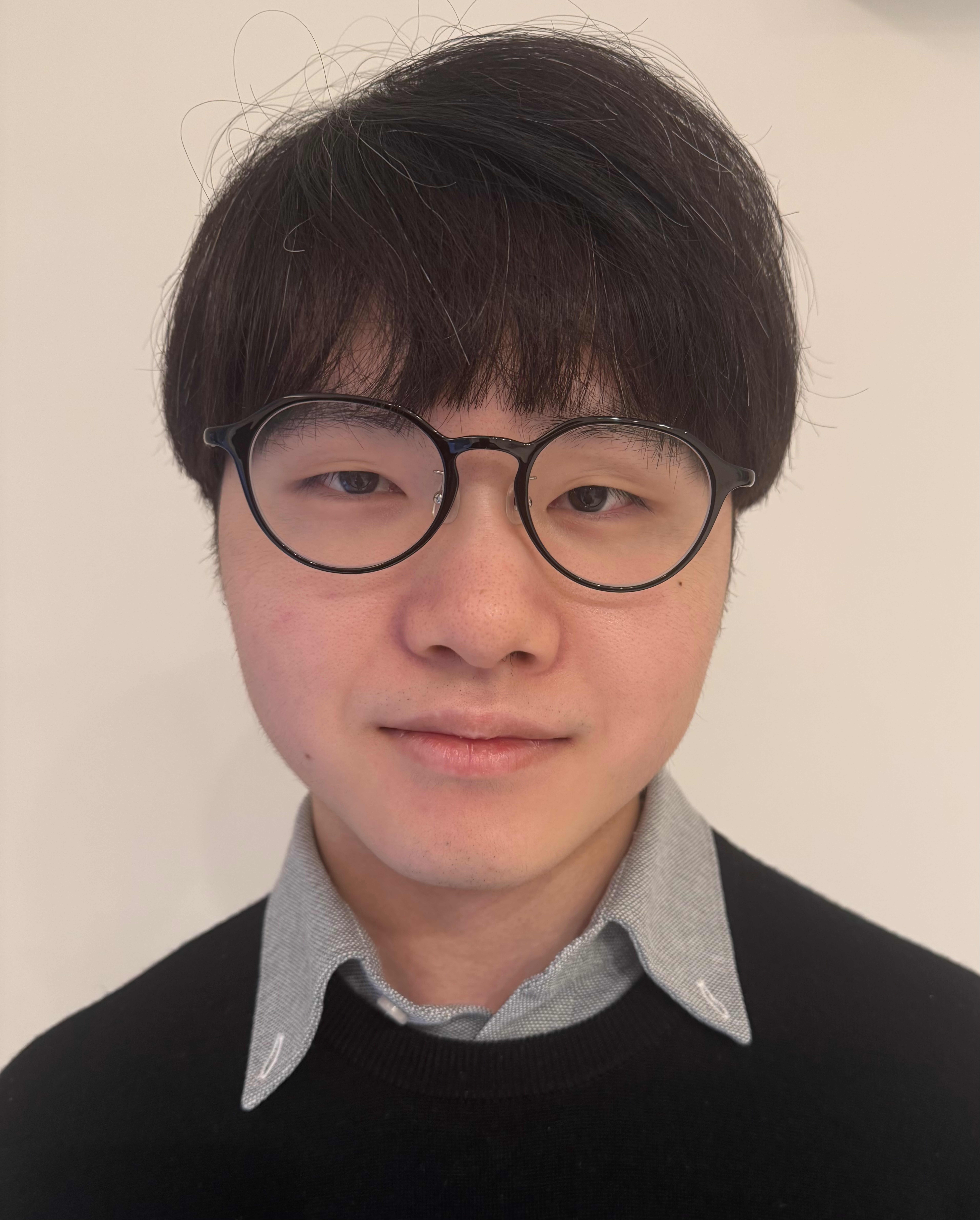}}]{Shunsei Yamagishi} received the B.S. degree and the M.S. degree in computer science and engineering from The University of  Aizu, Japan, in 2022 and 2024, respectively. He is currently working toward the Ph.D. degree in the Graduate School of Computer Science and Engineering, The University of Aizu, Japan. His research interests include algorithms for the attitude estimation for the Attitude and Heading Reference System, Pedestrian Dead Reckoning, signal processing, and sensor fusion methods.
\end{IEEEbiography}

\begin{IEEEbiography}[{\includegraphics[width=1in,height=1.25in,clip,keepaspectratio]{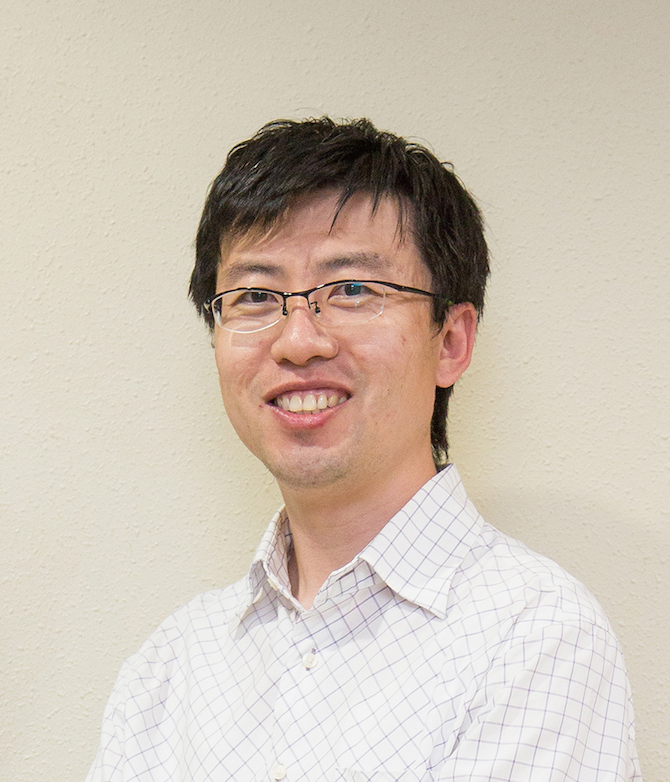}}]{Lei Jing} (M’12) received his Ph.D. degree in computer science and engineering from the University of Aizu, Japan, in 2008. He is currently a Senior Associate Professor at the School of Computer Science and Engineering, University of Aizu. His research interests include human position, posture, and motion tracking, soft circuit design, and the tactile internet. The applications of his work encompass human activity abnormality detection, sign language recognition, and human-robot interaction. He has published over 120 papers and holds six patents in related areas.
\end{IEEEbiography}

\vfill

\end{document}